\documentclass[twocolumn,aps,floatfix,superscriptaddress]{revtex4}

\usepackage{amsmath}
\usepackage{amssymb}
\usepackage{graphicx}
\usepackage{color}

\newcommand{\be}{\begin{equation}}
\newcommand{\ee}{\end{equation}}
\newcommand{\bea}{\begin{eqnarray}}
\newcommand{\eea}{\end{eqnarray}}
\newcommand{\bse}{\begin{subequations}}
\newcommand{\ese}{\end{subequations}}
\newcommand{\sma}{${\rm SrMn_2As_2}$}
\newcommand{\bma}{${\rm BaMn_2As_2}$}
\newcommand{\cma}{${\rm CaMn_2As_2}$}
\newcommand{\bms}{${\rm BaMn_2Sb_2}$}
\newcommand{\bmb}{${\rm BaMn_2Bi_2}$}
\newcommand{\sms}{${\rm SrMn_2Sb_2}$}
\newcommand{\smp}{${\rm SrMn_2P_2}$}
\newcommand{\bmp}{${\rm BaMn_2P_2}$}

\newcommand{\cas}{${\rm CaAl_2Si_2}$}
\newcommand{\tcs}{${\rm ThCr_2Si_2}$}

\begin{document}

\title{Antiferromagnetism in semiconducting ${\rm\bf SrMn_2Sb_2}$ and ${\rm\bf BaMn_2Sb_2}$ single crystals}

\author{N. S. Sangeetha}
\affiliation{Ames Laboratory, Iowa State University, Ames, Iowa 50011, USA}
\author{V. Smetana}
\author{A.-V. Mudring}
\affiliation{Ames Laboratory, Iowa State University, Ames, Iowa 50011, USA}
\affiliation{Department of Materials Science and Engineering, Iowa State University, Ames, Iowa 50011, USA}
\author{D. C. Johnston}
\affiliation{Ames Laboratory, Iowa State University, Ames, Iowa 50011, USA}
\affiliation{Department of Physics and Astronomy, Iowa State University, Ames, Iowa 50011, USA}

\date{\today}

\begin{abstract}

Crystals of \sms\ and \bms\ were grown using Sn flux and characterized by powder and single-crystal x-ray diffraction, respectively, and by single-crystal electrical resistivity~$\rho$, heat capacity~$C_{\rm p}$, and magnetic susceptibility~$\chi$ measurements versus temperature~$T$, and magnetization versus field $M(H)$ isotherm measurements. \sms\ adopts the trigonal \cas-type structure whereas \bms\ crystallizes in the tetragonal \tcs-type structure. The $\rho(T)$ data indicate semiconducting behaviors for both compounds with activation energies of $\geq0.35$~eV for \sms\ and 0.16~eV for \bms. The $\chi(T)$ and $C_{\rm p}(T)$ data reveal antiferromagnetic (AFM) ordering at $T\rm_N$ = 110~K for \sms\ and 450~K for \bms. The anisotropic $\chi (T\leq T\rm_N)$ data also show that the ordered moments in \sms\ are aligned in the hexagonal $ab$ plane whereas the ordered moments in \bms\ are aligned collinearly along the tetragonal $c$~axis. The $ab$-plane $M(H)$ data for \sms\ exhibit a continuous metamagnetic transition at low fields $0<H \lesssim 1$~T, whereas \bms\ exhibits no metamagnetic transitions up to 5.5~T\@.  The $\chi(T)$ data for both compounds and the $C_{\rm_p}(T)$ data for \sms\ and \bms\ indicate strong dynamic short-range AFM correlations above their respective $T\rm_N$ up to at least 900~K within a local-moment picture, corresponding to quasi-two-dimensional magnetic behavior. The present results and a survey of the literature for Mn pnictides with the \cas\ and \tcs\ crystal structures show that the $T_{\rm N}$ values for the \cas-type compounds are much smaller than those for the \tcs-type materials.

\end{abstract}

\maketitle

\section{Introduction}

The discoveries of high-$T_{\rm_c}$ superconductivity in layered iron pnictides and chalcogenides since 2008 revealed Fe-based materials to be a rich source of unconventional superconductors. In marked contrast to high-$T_{\rm_c}$ layered cuprates, the undoped ferropnictides are metallic and show itinerant antiferromagnetic (AFM) spin-density wave and coincident or nearly coincident structural transitions. The suppression of these transitions by chemical doping resulted in superconductivity in the ``1111''-type parent compound LaFeAsO with $T\rm_c = 26$~K in 2008 \cite{Kamihara}. Since then, many compounds with related layered crystal structures and chemical compositions were discovered which are  classifed as ``11''-type (e.g.\ binary iron chalcogenide FeSe), ``111''-type (e.g.\ ternary LiFeAs or NaFeAs), 1111-type with the primitive-tetragonal ZrCuSiAs-type structure with space group $P4/nmm$ (e.g.\ $R$FeAsO,  $R=$ rare earth), ``122''-type with the body-centered tetragonal \tcs\ structure with space group $I4/mmm$ (e.g.\ $A\rm Fe_2As_2$, $A=$ Ba, Sr, Ca, Eu), and other related structures \cite{{Just1996},{Johnston2010},{Stewart2011},{Scalapino2012}}. The highest reported $T\rm_c$ for a bulk Fe-based superconductor is 56~K for $\rm Gd_{0.8}Th_{0.2}FeAsO$ \cite{Wang2008}.

In attempts to further enhance the $T\rm_c$ of such superconductors and search for other exotic ground states, additional isostructural ccompounds based on other transition metals in place of Fe including Co, Mn, Cr, and Ni have been studied. These pnictides include metallic Co-based \cite{{Sefat2009},{Ohta2009}, {Cheng2012}, {Quirinale2013}, {Anand2014_CCA}, {Anand2014_BCA}, {Pandey2013}, {Jayasekara2013},{Sangeetha2016b}}, itinerant antiferromagnetic (AFM) Cr-based \cite{{Park2013}, {DJSingh}, {Filsinger}, {Naumov2017}, {Richard}, {Paramanik}, {Pfisterer1980}, {Pfisterer1983}, {Das2017b}}, semiconducting AFM Mn-based  \cite{{Brock1994},{Singh2009}, {Wang2011}, {Wang2009}, {Saparov2013},{Sangeetha2016},{McGuire2016}, {QZhang2016}, {Yanagi2009}} and superconducting Ni-based  \cite{{Li2014},{Tegel2008}, {Ronning2009}, {Bauer2008}} compounds.

Semiconducting layered quasi-two-dimensional AFM Mn pnictides received special attention as potential parent compounds for superconductivity because they could form a bridge between the superconducting iron-arsenide and cuprate families of high-$T\rm_c$ materials. The 1111-type LaMnPO and 122-type \bma\ have been most extensively studied. These compounds exhibit insulator (at $T=0$)  to metal transitions either upon doping or application of pressure, but no superconductivity in these compounds has been observed.  The AFM ordering (N\'eel) temperature~$T_{\rm N}$, ordered moment $\mu$ at low tmperatures~$T$, and activation energy~$\Delta$ have been reported for several semiconducting 1111-type Mn pnictides, including \mbox{LaMnPO}~\cite{Yanagi2009}, LaMnAsO  \cite{McGuire2016}, and LaMnSbO \cite{QZhang2016}, as summarized in Table~\ref{Table:MnPnMagProps} for these compounds and also for those below.  They tend to have a simple checkerboard (G-type) AFM structure, indicating stronger nearest-neighbour (NN) exchange interactions~$J_1$ compared to twice the next-nearest-neighbor interactions~$J_2$, i.e., $J_1>2J_2$ \cite{Johnston2010}. A high-pressure measurement on LaMnPO showed a semiconductor to metal transition at a pressure of 10~GPa \cite{Simonson2012a}. Fluorine-doped LaMnPO$_{1-x}$F$_x$ also exhibits an insulator to metal transition on doping which only weakly affects $T\rm_N$ and $\mu$ \cite{Simonson2011}, similar to La$_{1-x}$Sr$_x$MnAsO \cite{Sun2012}. On the other hand, hydrogen- and deuterium-substituted LaMnAsO$_{1-x}$(H/D)$_x$ exhibits insulator to metal transitions associated with the emergence of itinerant ferromagnetism \cite{Hanna2013}.

The 122-type BaMn$_2X_2$ compounds with the ThCr$_2$Si$_2$-type structure containing $X$ = P, As, or Bi have also been extensively studied. \bmp\ \cite{Brock1994}, \bma\ \cite{Singh2009,Singh2009b}, and \bmb\ \cite{{Saparov2013},{Calder2014}} undergo G-type AFM ordering with the ordered moments aligned along the $c$~axis, where the NN exchange interactions $J_1$ are again dominant. A high-pressure study of \bma\ showed an insulator to metal transition at 5.8~GPa with a downturn in the resistivity below $\sim 17$~K and no change in the crystal structure \cite{Satya2011}. Interestingly, substitution of only 1.6\% of K for Ba results in a metallic ground state in \bma\@ \cite{Pandey2012}. Furthermore, itinerant FM occurs in Ba$_{1-x}$K$_x$Mn$_2$As$_2$ with $x=0.19$ and 0.26 \cite{Bao2012} and half-metallic FM behavior is found below the Curie temperature $T\rm_C \approx 100$~K for 40\% K-doped \cite{Pandey2013b} and 60\% Rb-doped \bma\@ \cite{Pandey2015}. This itinerant FM \cite{Ueland2015} is aligned in the $ab$~plane and coexists with the G-type local-moment AFM of the Mn spins aligned along the $c$~axis \cite{Lamsal2013}.  Similar to \bma\@, the isostructural AFM semiconductor BaMn$_2$Bi$_2$ exhibits metallic behavior with K doping and the magnetic character of Ba$_{1-x}$K$_x$Mn$_2$Bi$_2$ up to 36\% K doping is inferred to be local-moment AFM~\cite{Saparov2013}.

Interestingly, unlike body-centered tetragonal 122-type BaMn$_2$(P, As, Sb, Bi)$_2$,  the 122-type (Ca, Sr)Mn$_2$(P, As, Sb, Bi)$_2$ compounds crystallize in the trigonal \cas-type structure with space group $P\bar{3}m1$.   Empirically, the \cas-type structure is found if the transition metal has a $d^0$, d$^5$ or $d^{10}$ electronic configuration whereas the \tcs-type structure has no such preferences. The Mn sublattice can be viewed either as a corrugated honeycomb layer or a triangular-lattice  bilayer which suggests the possibility of geometric frustration. Some of these trigonal compounds are known to be AFM semiconductors: \smp\ \cite{Brock1994}, \cma\ \cite{Sangeetha2016}, \sma\ \cite{{Sangeetha2016},{Das2017}}, CaMn$_2$Sb$_2$ \cite{{Bridges2009},{Ratcliff2009},{Simonson2012}}, and CaMn$_2$Bi$_2$ \cite{Gibson2015}.  

The compounds in Table~\ref{Table:MnPnMagProps} also manifest strong dynamic short-range AFM correlations at $T\gg T\rm_N$, similar to \bma\ \cite{Johnston2011}. Neutron diffraction studies of both CaMn$_2$Sb$_2$ and CaMn$_2$Bi$_2$ revealed a collinear AFM structure with the ordered moments aligned in the $ab$~plane \cite{{Gibson2015},{McNally2015}}.  However, whereas one neutron structural study of CaMn$_2$Sb$_2$ concluded that collinear AFM moments are aligned in the $ab$~plane \cite{Ratcliff2009}, another suggested that the collinear ordered moments are canted by $25^\circ$ with respect to the $ab$~plane \cite{Bridges2009}. Single-crystal neutron diffraction measurements on trigonal \sma\ showed that the Mn moments are ordered in the $ab$~plane in a collinear N\'eel AFM structure below $T_{\rm N}=118(2)$~K \cite{Das2017} and magnetic susceptibility data exhibited quasi-2D behavior at $T\gg T_{\rm N}$ \cite{Sangeetha2016}, as does \tcs-type \bma\ as noted above.  \smp\ exhibits a novel pressure-induced structural transition from the trigonal \cas-type to the body-centered tetragonal \tcs-type structure \cite{Xie2017}.

\begin{table}
\caption{\label{Table:MnPnMagProps} Properties of 1111-type and 122-type Mn pnictides, including results from the present work (PW).  Included are the structure type, AFM ordering temperatures~$T_{\rm N}$, low-temperature ordered moment~$\mu$, and semiconducting activation energy~$\Delta$.  The ZrCuSiAs structure is primitive tetragonal with space group $P4/nmm$, the \tcs\ structure is body-centered tetragonal with space group $I4/mmm$, and the \cas\ structure is trigonal with space group $P\bar{3}m1$.  Unless otherwise indicated, $\Delta$ was determined from electrical resistivity data with the current in the crystallographic $ab$~plane of single-crystal samples.}
\begin{ruledtabular}
\begin{tabular}{lccccc}
Compound 	& Structure		&	$T_{\rm N}$	& 	$\mu$			& 	$\Delta$  &	Ref.		\\
		&	Type			&	(K)			&	($\mu_{\rm B}$/Mn)	&	(eV)		&			\\
\hline 

\underline{1111-type}\\
LaMnPO	&  	ZrCuSiAs 		& 	375  		&	2.26				&			&	\cite{Yanagi2009}	\\
LaMnAsO	&	ZrCuSiAs		& 	360  		&	3.33				&			&	\cite{McGuire2016}	\\
LaMnSbO	&  	ZrCuSiAs 		& 	255			& 	3.45				& 			&	\cite{QZhang2016}	\\
\hline
\underline{122-type}\\
\bmp\	&	\tcs\	&				&	4.2				&	0.073	& 	\cite{Brock1994}\footnotemark[1]	\\
\bma\	&	\tcs\	&	625			&	3.88				&	0.027	&	\cite{Singh2009,Singh2009b}	\\
\bms\	&	\tcs\	&				&					&	1.12\footnotemark[2]		&	\cite{Wang2009}	\\
\bms\	&	\tcs\	&	450			&					&	0.16		&	PW	\\
\bmb\	&	\tcs\	&	400			&	3.83				&	0.003	&	\cite{Saparov2013,Calder2014}	\\
\hline
\cma\	&	\cas\		&	62			&					&	0.061	&	\cite{Sangeetha2016}	\\
CaMn$_2$Sb$_2$	&	\cas\	&	85--88			&	3.46				&			&	\cite{{Bridges2009},{Ratcliff2009},{Simonson2012}} \\
CaMn$_2$Bi$_2$	&	\cas\	&	150			&	3.85				&	0.031	&	\cite{Gibson2015} \\
\smp\	&	\cas\		&	53			&					&	0.013	&	\cite{Brock1994}\footnotemark[1]	\\
\sma\	&	\cas\		&	120			&	3.6				&	0.085	&	\cite{{Sangeetha2016},{Das2017}}	\\
\sms\	&	\cas\		&	110			&					&	$\geq0.35$	&	PW		\\

\end{tabular}
\end{ruledtabular}
\footnotetext[1]{Polycrystalline sample.}
\footnotetext[2]{Measured at 470--773~K.}
\end{table}
 
In order to extend the above work on the Mn pnictides and attempt to discover new properties, herein we report the growth, crystal structure, $\rho$(T), magnetization versus applied magnetic field $M(H)$ isotherms, $\chi(T)$ and heat capacity $C_{\rm p}(T)$ measurements of \bms, which crystallizes in the \tcs-type structure \cite{Brechtel1979}, and \sms\ which has the \cas-type structure \cite{Cordier1976}.  Since the Mn sublattice in \sms\ is a corrugated honeycomb or triangular bilayer sublattice whereas  the Mn sublattice in \bms\ is a stacked square lattice, the difference in the geometry of Mn network between \bms\ and \sms\ may be important to take into account when comparing their properties.

Previous theoretical electronic structure studies on \bms\ suggested a G-type AFM ordering \cite{{Xia2008},{An2009}}, but $T_{\rm N}$ and the AFM structure have not been experimentally determined.  Transport measurements revealed that \bms\ has a room-temperature Seebeck coefficient of $\sim 225~\mu$V/K, consistent with semiconducting behavior. With regard to \sms, a polycrystalline sample showed ``complex magnetic ordering of the Mn moments below $\approx 265$~K'' and $\rho(T)$ measurements on a crystal suggested a ``metallic-like temperature dependence'' \cite{Bobev2006}.  However, in contradiction to both of these results, the present study on \sms\ crystals reveals AFM ordering at $T\rm_N$=120~K with an insulating ground state.  

The remainder of the paper is organized as follows.  Following the experimental details in Sec.~\ref{ExpDetails},  $\rho(T)$ data for \bms\ and \sms\ crystals are presented in Sec.~\ref{Sec:Rho}. Our studies of $M(H)$ and $\chi(T)$ for these crystals are presented in Sec.~\ref{Sec:MandChi}.  We suggest in Secs.~\ref{Sec:Rho} and~\ref{Sec:SrMn2Sb2MHT} the probable origins of the erroneous experimental $\rho(T)$ and $\chi(T)$ results reported in Ref.~\cite{Bobev2006}, respectively.  The heat capacity $C_{\rm p}(T)$ measurements on \sms\ and \bms\  are presented in Sec.~\ref{Sec:HC}, and a summary is given in Sec.~\ref{Sec:Summary}.

\section{\label{ExpDetails} Experimental Details}

Single crystals of \bms\ and \sms\ were grown in Sn flux. High-purity elements Sr (99.99\%) and Ba (99.99\%) from Sigma Aldrich, and Mn (99.95\%), Sb (99.9999\%), and Sn (99.999\%) from Alfa Aesar, were weighed out in the molar ratio (Ba, Sr):Mn:Sb:Sn = 1.05:2:2:20 and placed in an alumina crucible that was sealed under an Ar pressure of $\approx 1/4$ atm in a silica tube. Excess Sr or Ba was used in the synthesis to avoid the formation of MnSb and thus to assure the formation of magnetically pure samples. After preheating the mixture at 600~$^{\circ}$C for 7~h, the assembly was heated to 1150~$^{\circ}$C at a rate of 50~$^{\circ}$C/h and held at this temperature for 15~h for homogenization. Then the furnace was slowly cooled at a rate of 4~$^{\circ}$C/h to 700~$^{\circ}$C\@. Shiny platelike and  hexagon-shaped single crystals of \bms\ and \sms\, respectively, were obtained after decanting the Sn flux using a centrifuge.  Subsequent handling of the crystals was carried out in a dry box and exposure of them to air during transfer to measurement instruments was kept to a minimum.

Semiquantitative chemical analyses of the single crystals were performed using a JEOL scanning electron microscope (SEM) equipped with an EDX (energy-dispersive x-ray analysis) detector, where a counting time of 120 s was used. A room-temperature powder x-ray diffraction (XRD) pattern was recorded on crushed single crystals of \sms\ using a Rigaku Geigerflex powder diffractometer with Cu K$\alpha$ radiation ($\lambda = 1.5418$~\AA) at diffraction angles 2$\theta$ from 10$^{\circ}$ to 110$^{\circ}$ with a 0.02$^{\circ}$ step width.  The data were analysed by Rietveld refinement using FullProf software \cite{fullprof}. 

Single-crystal x-ray structural analysis of \bms\ was performed at room temperature using a Bruker D8 Venture diffractometer equipped with a Photon 100 CMOS detector, a flat graphite monochromator and a Mo~K$\alpha$ I$\mu$S microfocus x-ray source ($\lambda = 0.71073$~\AA) operating at a voltage of 50~kV and a current of 1~mA\@. The raw frame data were collected using the Bruker APEX3 program \cite{APEX2015}, while the frames were integrated with the Bruker SAINT software package \cite{SAINT2015} using a narrow-frame algorithm integration of the data and were corrected for absorption effects using the multiscan method (SADABS) \cite{Krause2015}.  The atomic thermal factors were refined anisotropically.  Initial models of the crystal structures were first obtained with the program SHELXT-2014 \cite{Sheldrick2015A} and refined using the program SHELXL-2014 \cite{Sheldrick2015C} within the APEX3 software package.

Magnetization~$M$ versus temperature~$T$ measurements at fixed field over the $T$ range ${\rm 1.8 \leq}~T~{\rm \leq 350~K}$ and $M$ versus applied field~$H$ isotherm measurements for $H \leq 5.5$~T were carried out using a Quantum Design, Inc., Magnetic Properties Measurement System (MPMS). The high-temperature $M(T)$ for ${\rm 300~K \leq T \leq 900~K}$ was measured  using the vibrating sample magnetometer (VSM) option of a Quantum Design, Inc., Physical Properties Measurement System (PPMS)\@.  Four-probe dc $\rho(T)$ and $C{\rm_p}(T)$ measurements were carried out on the PPMS, where electrical contacts to a crystal for the $\rho(T)$ measurements were made using annealed 0.05~mm diameter Pt wires and silver epoxy.

\section{Experimental Results}

\subsection{\label{Sec:Struct} Crystal Structure}

\begin{figure}
\includegraphics[width=3.4in]{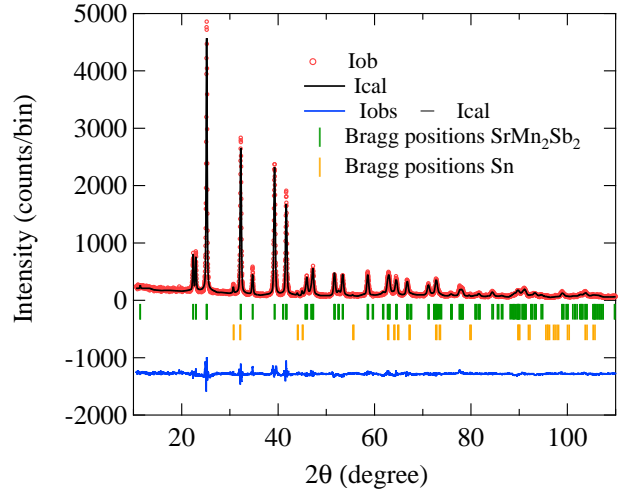}
\caption{(Colour online) Powder x-ray diffraction patterns (open circles) of \sms\ at room temperature. The solid line represents the Rietveld refinement fit calculated for the \cas-type trigonal structure with space group P$\bar{3}$m1 together with small amounts of  adventitious elemental Sn impurity on the crystal surfaces from the Sn flux.}
\label{Fig_xrd}
\end{figure}

\begin{table}
\caption{\label{Table_SXRD} Refined crystallographic parameters obtained from XRD of \sms\ and \bms\ crystals. The atomic coordinates in \sms\ in hexagonal notation are Sr: 1a (0, 0, 0); Mn: 2d (1/3, 2/3, $z \rm_{Mn}$); and Sb: 2d (1/3, 2/3, $z{\rm_{Sb}}$). The atomic coordinates in \bms\ are Ba: 1a (0, 0, 0); Mn: 2d (0, 1/2, 1/4); and Sb: 2d (0, 0, $z{\rm_{Sb}}$).  The shortest Mn-Mn interatomic distances in \sms\ and \bms\  [see Fig.~\ref{Fig_unitcell}] are also listed. }

\begin{tabular}{p{3.5cm}|p{2.5cm}|p{2.3cm}}
\hline
\hline 
\\
 & \sms\  & \bms\ \\
\hline
Structure  & \cas-type trigonal &   \tcs-type tetragonal \\

Space group  & P$\bar{3}$m1  & I$4/mmm$ \\
\\
Lattice parameters &  &   \\
\hspace{0.8cm} $a$ (\AA )  & 4.5888(2)  &   4.397(4)\\
\hspace{0.8cm} $c$ (\AA )  & 7.7529(4)  &  14.33(2)\\
\hspace{0.8cm} $c/a$  	  & 1.6895(2)  &  3.259(8)\\
\hspace{0.8cm} $V_{{\rm cell}}$ (\AA $^{3}$)  & 141.38(1)  &  277.15(6) \\
Atomic coordinates & & \\
\hspace{0.8cm} $z{\rm_{Mn}}$  & 0.62123(12)  &  \\
\hspace{0.8cm} $z{\rm_{Sb}}$   & 0.26133(20) &  0.3642(1)  \\
\hline
Shortest Mn-Mn \\
distances (\AA) &  &   \\
\hspace{0.8cm} $d_1$  & 3.2471(7)  &  3.2140(3) \\
\hspace{0.8cm} $d_2$   & 4.5965(3) &   4.4180(6) \\
\hspace{0.8cm} $d_3$   & 5.6277(5) &  6.2480(6) \\
\hspace{0.8cm} $d_{z1}$   & 6.4753(12) &  7.2010(2) \\
\hspace{0.8cm} $d_{z2}$   & 7.778(2) & 14.4020(3)  \\
\hline
\hline
\end{tabular}
\end{table}

SEM images of the crystal surfaces indicated single-phase crystals. EDX analyses of the chemical compositions were in agreement with the expected 1:2:2 stoichiometry of the compounds and the amount of Sn incorporated into the crystal structure from the Sn flux is zero to within experimental error.

Powder XRD data on crushed \sms\ single crystals confirmed their single-phase nature. The Rietveld refinement of the XRD pattern is shown in Fig.~\ref{Fig_xrd}, confirming that \sms\ has the trigonal \cas-type structure.  Weak peaks from adventitious Sn from the flux are also visible and were refined together with the main phase.  The crystallographic parameters of \sms\ are listed in Table~\ref{Table_SXRD}. The hexagonal lattice parameters $a$ and $c$ obtained are in good agreement with previously reported values \cite{Cordier1976,Bobev2006}. 

Single-crystal XRD measurements on \bms\ confirmed the single-phase nature of the compound and the \tcs-type crystal structure.  The crystallographic parameters are listed in Table~\ref{Table_SXRD}. The lattice parameters $a$ and $c$ are in good agreement with the previous values~\cite{Brechtel1979}.

\begin{figure}
\includegraphics[width=3.3in]{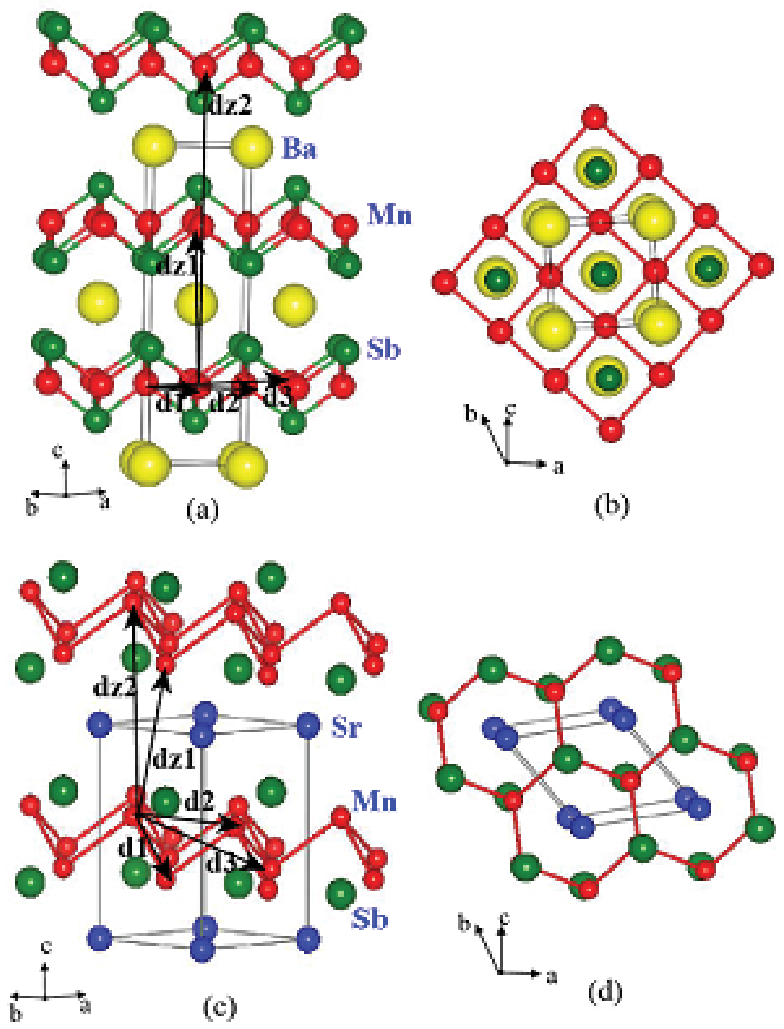}
\caption{(Colour online) Outlines of the unit cells of (a)~\bms\ ($I4/mmm$, \tcs-type) and (c)~\sms\ (P$\bar{3}$m1, \cas-type). The smallest Mn-Mn interatomic distances within the Mn layer ($d_1$, $d_2$ and $d_3$) and between layers ($d_{z1}$ and $d_{z2}$) are indicated by arrows. The projections of the manganese networks onto the $ab$~plane with a slight $c$-axis tilt are shown in (b)~\bms\ (square net) and (d)~\sms\ (corrugated honeycomb net or triangular bilayer).}
\label{Fig_unitcell}
\end{figure}

Figures~\ref{Fig_unitcell}(a) and~\ref{Fig_unitcell}(c) show perspective views of the unit cells of \bms\ (tetragonal \tcs-type) and \sms\ (trigonal \cas-type), respectively. Both structures look similar in the way they are composed of Mn-Sb tetrahedra separated by Sr/Ba layers. The primary difference between them is the geometry of the Mn layers.  In \bms, the Mn network is square planar where each Mn atom is coordinated by four Mn atoms at a 90$^\circ$ angle between them as shown in Fig.~\ref{Fig_unitcell}(b). In \sms,\ the planar Mn network can be considered to be either a corrugated Mn honeycomb layer or a double triangular-lattice layer as shown in Fig.~\ref{Fig_unitcell}(d). The difference between the structures of the Mn layers in the two compounds may be expected to have a significant influence on their physical properties.

The three smallest Mn-Mn intralayer distances $d_1$, $d_2$, and $d_3$ and the smallest Mn-Mn interlayer distances $d_{z1}$ and $d_{z2}$ are shown by arrows in Figs.~\ref{Fig_unitcell}(a) and~\ref{Fig_unitcell}(c) for \bms\ and \sms\, respectively. These distances are listed in Table~\ref{Table_SXRD}.

\subsection{\label{Sec:Rho}  In-Plane Electrical Resistivity}

\begin{figure}
\includegraphics[width=3.3in]{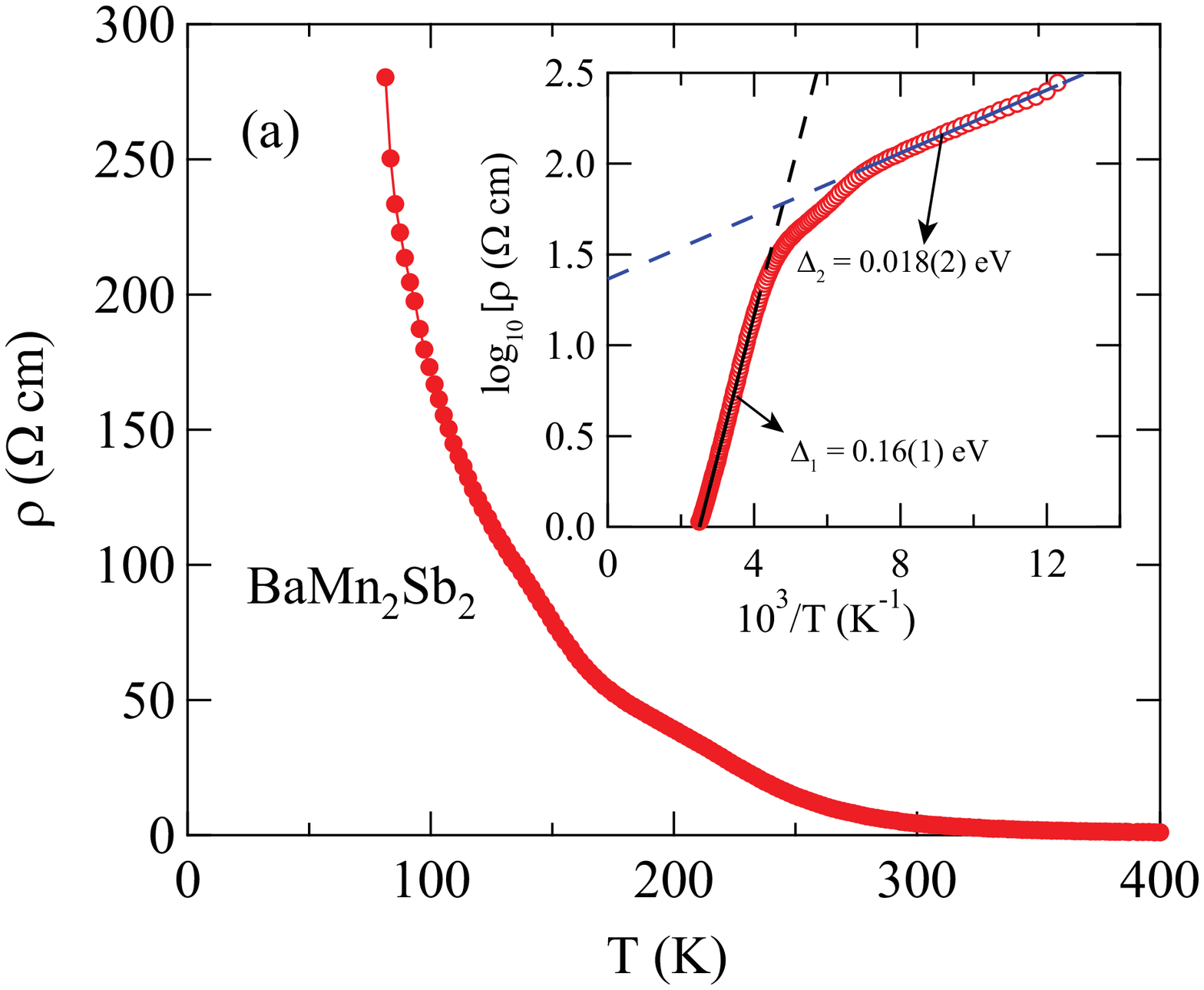}
\includegraphics[width=3.3in]{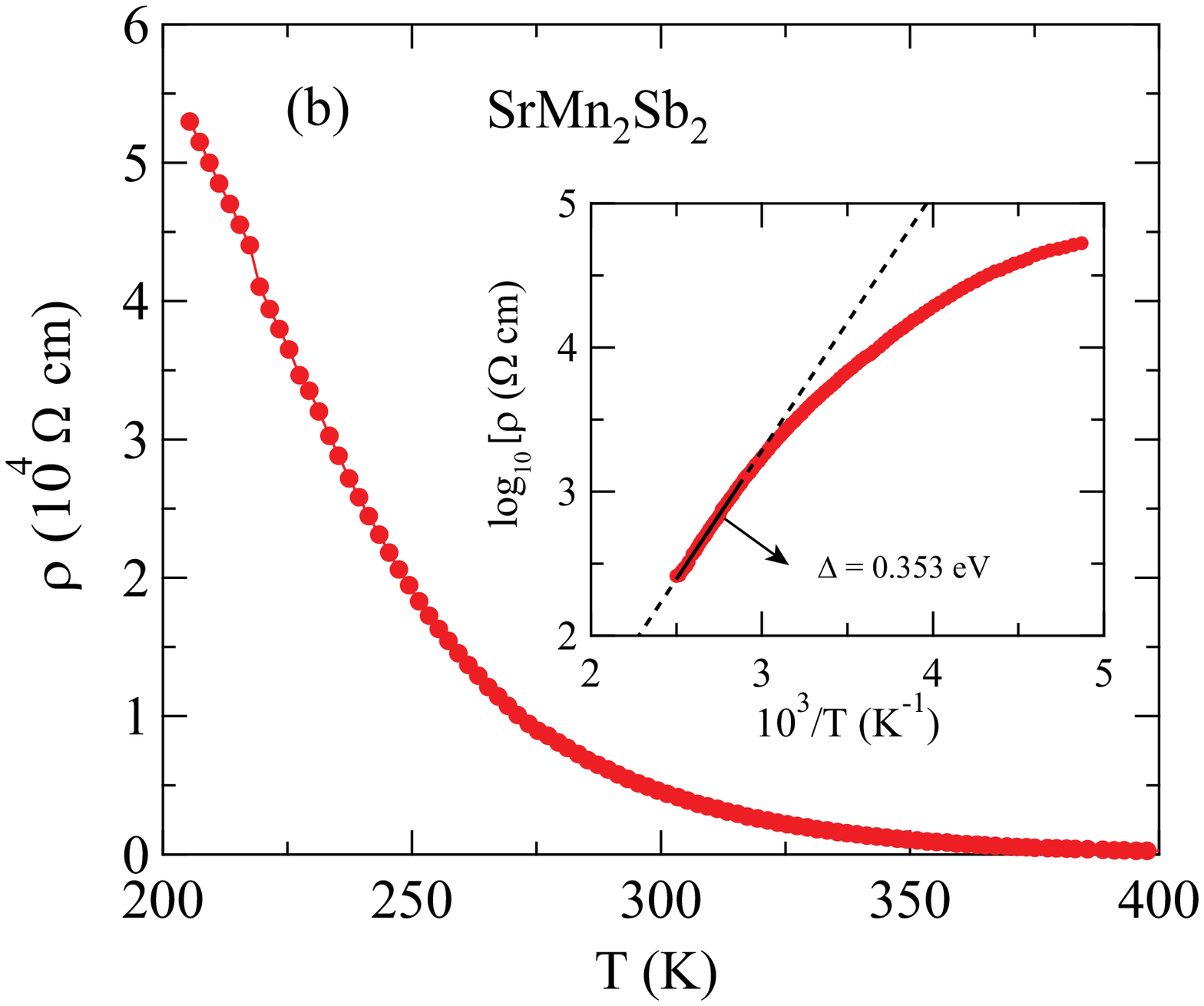}
\caption{(Colour online) Temperature $T$ dependence of the electrical resistivity, $\rho(T)$, in the $ab$~plane for (a)~\bms\ and (b)~\sms. The insets show plots of $\log_{10} \rho$ versus $1/T$\@. The solid straight lines through the inset data are fits over restricted temperature intervals by Eq.~(\ref{rho_fit}) as discussed in the text, and the dashed lines are extrapolations.}
\label{Fig_rho_BaSrMn2Sb2}
\end{figure}

Figures~\ref{Fig_rho_BaSrMn2Sb2}(a) and \ref{Fig_rho_BaSrMn2Sb2}(b) show the $ab$-plane $\rho(T)$ for \bms\ and \sms, respectively.  For both compounds, $\rho$ first increases slowly with decreasing~$T$ below 400~K, but then $\rho$ increases rapidly with decreasing~$T$ below 300~K\@.   The data clearly indicate that both \bms\ and \sms\ have an insulating ground state. We fitted $\rho(T)$ data over restricted temperature intervals by
\be
\log_{10}\rho = A + 0.4343\left(\frac{\Delta}{k{\rm_B}T}\right),
\label{rho_fit}
\ee
where $A$ is a constant, $k\rm_B$ is Boltzmann's constant and $\Delta$ is the activation energy.

Plots of $\log_{10} \rho$ vs $1/T$  are shown in the insets of Figs.~\ref{Fig_rho_BaSrMn2Sb2}(a) and~\ref{Fig_rho_BaSrMn2Sb2}(b) for \bms\ and \sms,\ respectively. For \bms,\ the data between  1)~240~K and 400~K and between 2)~80~K and 140~K are nearly linear in $T$ and were separately fitted by Eq.~(\ref{rho_fit}), yielding the intrinsic and extrinsic activation energies $\Delta_1 =  0.16$~eV and $\Delta_2 = 0.018$~eV, respectively, corroborating previous results \cite{Wang2009}. The fits are shown as the solid straight lines in the inset of Fig.~\ref{Fig_rho_BaSrMn2Sb2}(a), and the extrapolations as dashed lines.   The fitted activation energies are listed in the inset and Table~\ref{Table:MnPnMagProps} and are of the same order as found previously for isostructural \bmp\ \cite{Brock1994} and \bma\ \cite{Singh2009}.

In the case of \sms, there was no extended region of the $\ln\rho$ versus~$1/T$ plot that showed linear behavior, so the data between 340~K and 400~K were fitted by Eq~(\ref{rho_fit}), yielding the lower limit to the intrinsic activation energy as $\Delta = 0.35$~eV as shown in the inset and Table~\ref{Table:MnPnMagProps}. The fit is shown as the solid straight line in the inset of Fig.~\ref{Fig_rho_BaSrMn2Sb2}(b) and the extrapolations by dashed lines. 

Our results for \sms\ contradict a previous report \cite{Bobev2006} where it was claimed that \sms\ exhibits a ``metallic-like'' in-plane $\rho(T)$.  In particular, the $\rho(T)$ in Fig.~6 of Ref.~\cite{Bobev2006} increases almost linearly from $\approx 30$\,m$\Omega$\,cm at $T\to0$ to $\approx 100$\,m$\Omega$\,cm at $T=300$~K\@.  These values are too large for a metallic layered pnictide \cite{Johnston2010}.  Furthermore, the  intrinsic values in Fig.~\ref{Fig_rho_BaSrMn2Sb2}(b) over the same $T$ range are orders of magnitude larger than in Ref.~\cite{Bobev2006}.  We therefore suggest that the origin of these discrepancies is the presence of metallic Sn inclusions in the crystal measured in Ref.~\cite{Bobev2006} which would short-circuit the much higher intrinsic resistance of the crystal and lead to their observed $T$ dependence of $\rho(T)$.  We note that the lowest temperature of the $\rho(T)$ measurement in the inset to Fig.~6 of Ref.~\cite{Bobev2006} was 4.5~K, just above the superconducting $T_{\rm c}=3.7$~K of Sn metal which would likely have been detected by extending their $\rho(T)$ measurement to somewhat lower~$T$\@.

\subsection{\label{Sec:MandChi} Magnetization and Magnetic Susceptibility}

\subsubsection{\label{Sec:SrMn2Sb2MHT} \sms}

\begin{figure}
\includegraphics[width=3.3in]{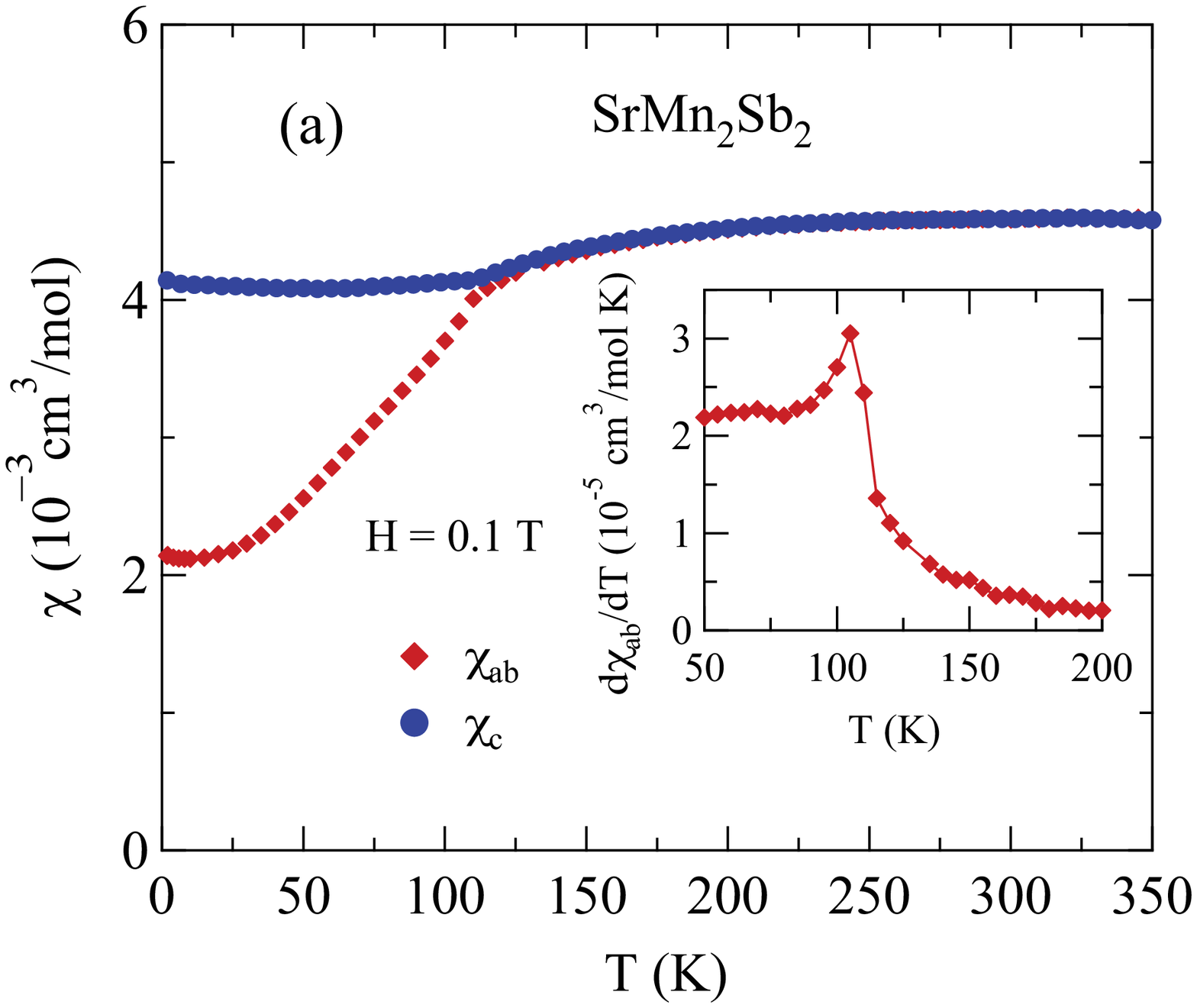}
\includegraphics[width=3.3in]{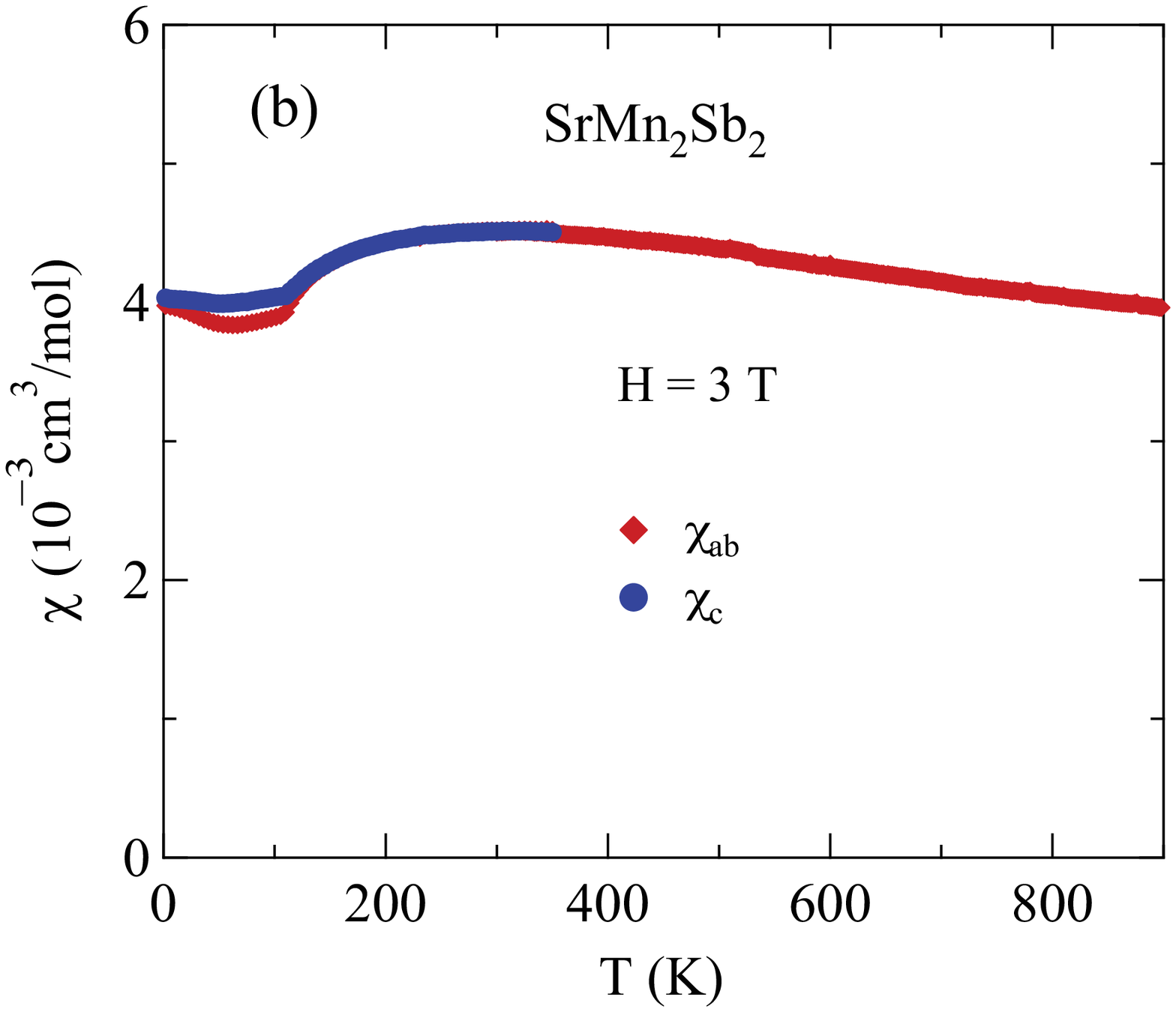}
\caption{(Color online) (a)~Zero-field-cooled magnetic susceptibility $\chi$ of \sms\ versus temperature $T$ in a magnetic field $H=0.1$~T applied in the $ab$~plane ($\chi_{ab}$) and along the $c$~axis ($\chi_c$). Inset: Derivative $d\chi(T)/dT$ versus $T$ for $H\,||\,ab$ to identify $T_{\rm N}$.  (b)~$\chi_{ab}$ and $\chi_c$ versus $T$ for $1.8~{\rm K}\leq T \leq$ 900~K measured in $H=3$~T\@. }
\label{Fig_SrMn2Sb2_MT}
\end{figure}

\begin{figure}
\includegraphics[width=3.3in]{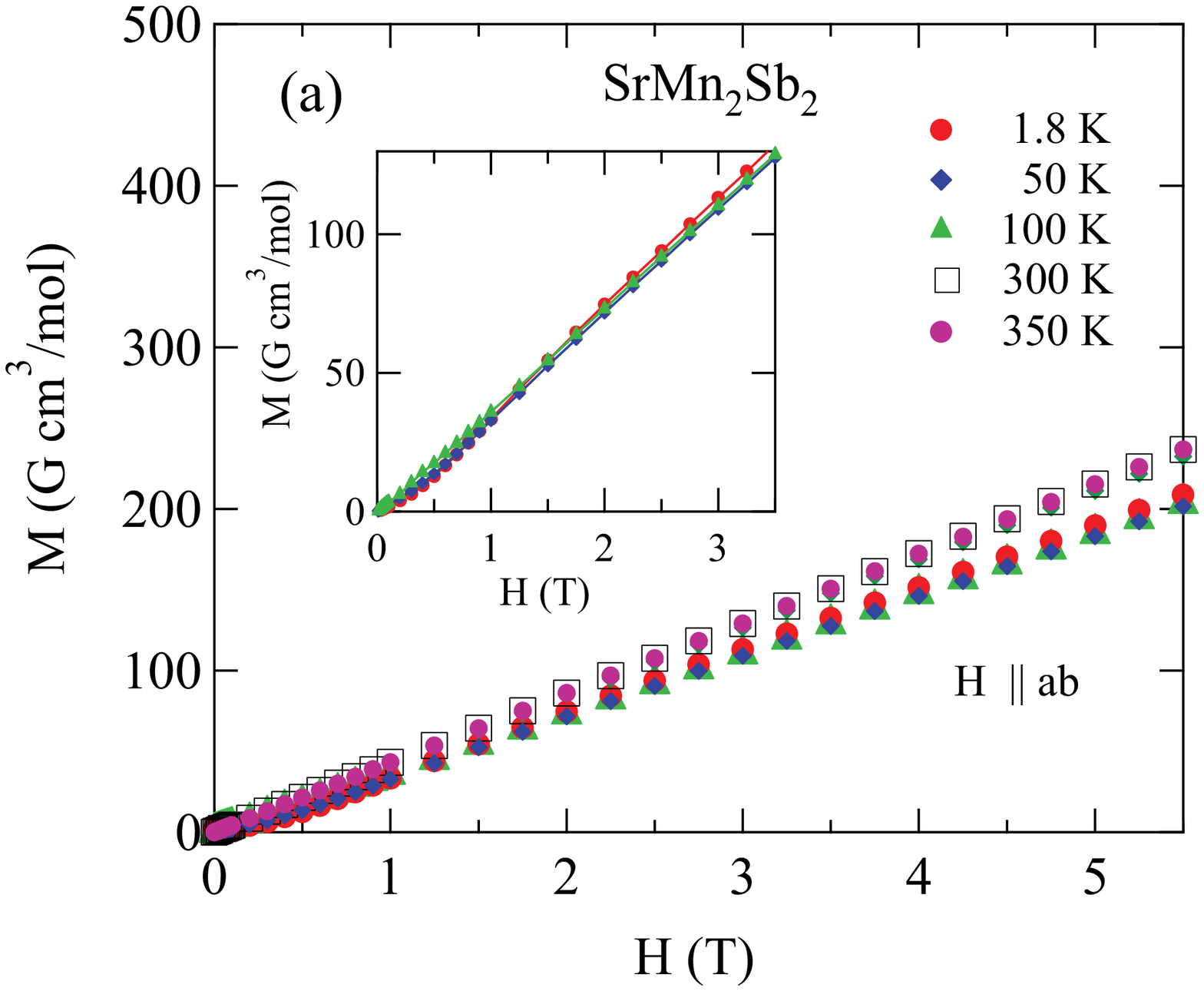}
\includegraphics[width=3.3in]{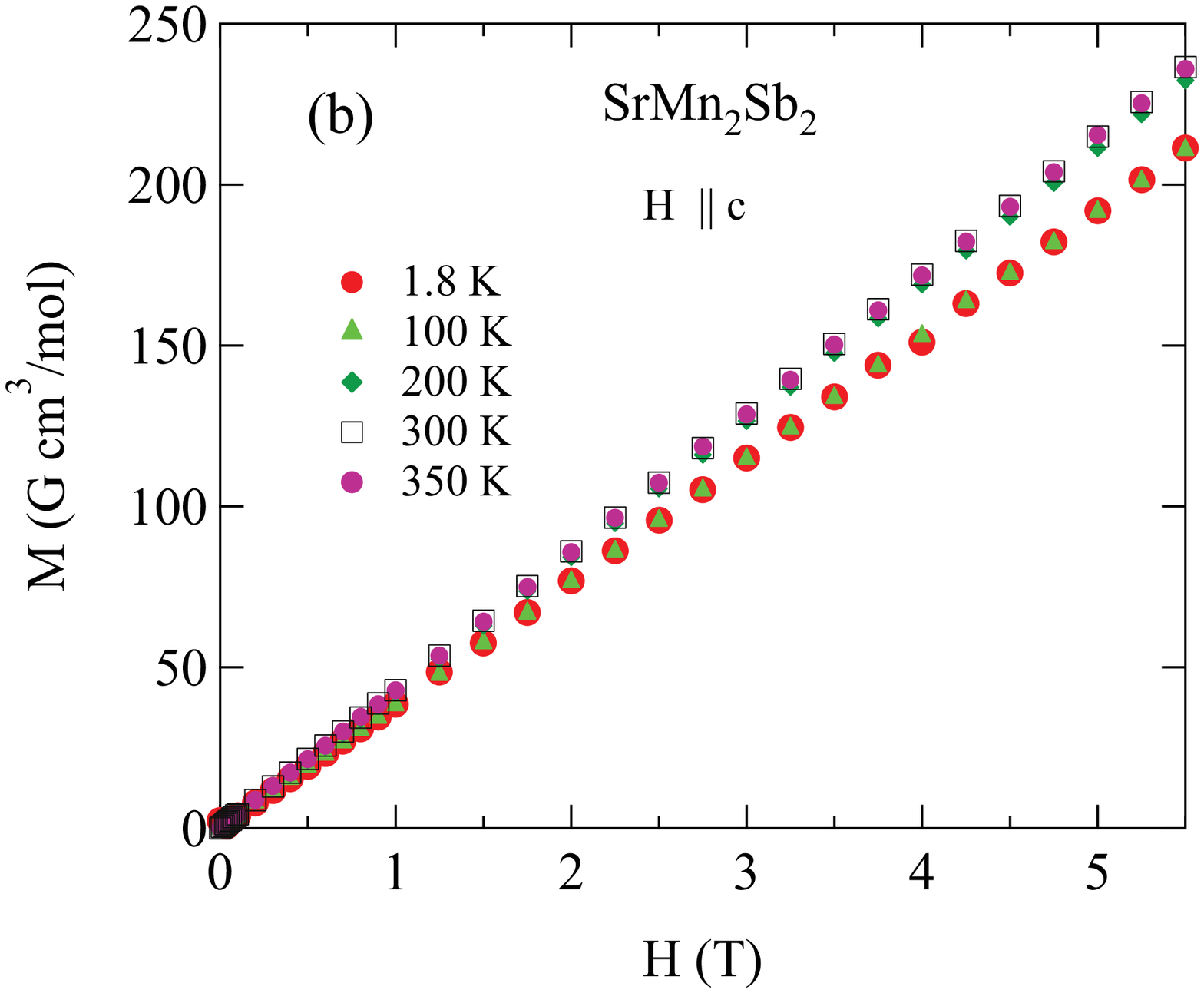}
 \protect\caption{(Color online) Magnetization $M$ of \sms\ as a function of magnetic field $H$ at various temperatures $T$ with (a) $H$ in the $ab$~plane ($H\,||\, ab$) and (b) $H$ along $c$ axis ($H\,||\, c$).   The inset in~(a) is an expanded plot of $M(H)$.}
\label{Fig:SrMn2Sb2_MH}
\end{figure}

\begin{figure}
\includegraphics[width=3.3in]{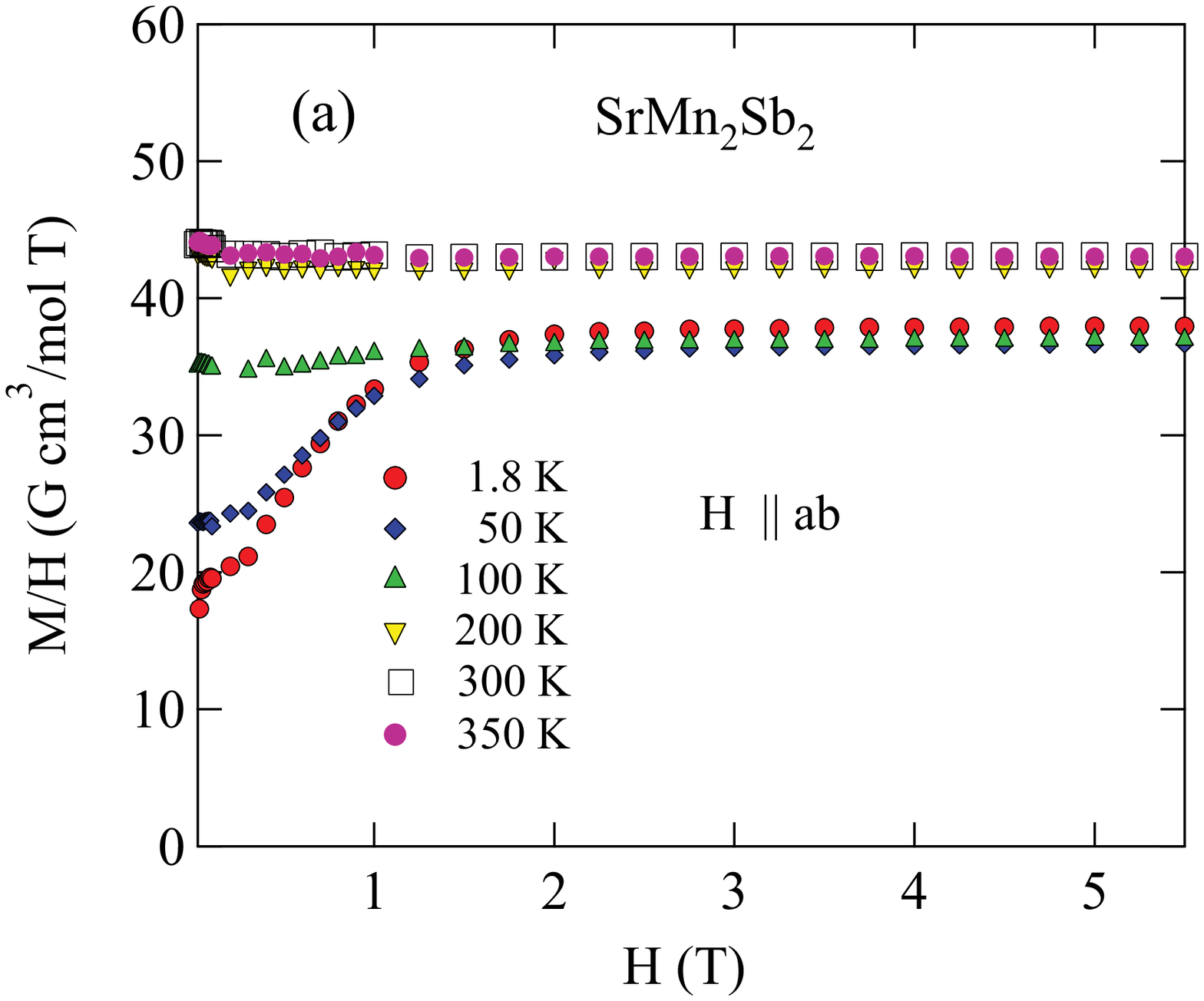}
\includegraphics[width=3.3in]{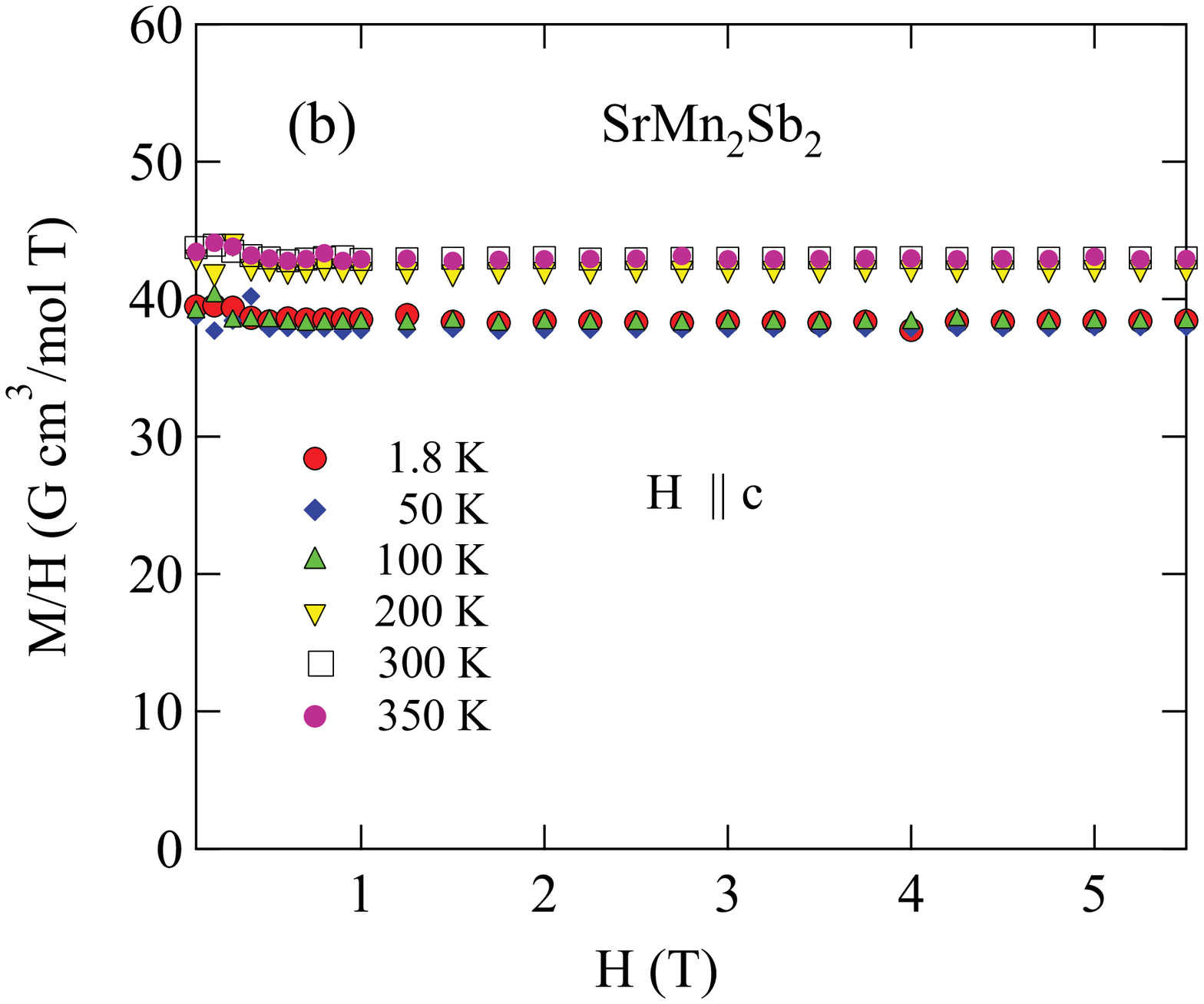}
 \protect\caption{(Color online) $M/H$ versus $H$ of \sms\ at various listed temperatures $T$ with (a)~$H$ in the $ab$~plane ($H\,||\, ab$) and (b)~$H$ along the $c$ axis ($H\,||\, c$). The data in~(a) for $T=1.8$~K and 50~K show a smooth metamagnetic transition that is spread out between $H=0$ and $H=1$~T.}
\label{Fig:SrMn2Sb2_MonH}
\end{figure}

Figure~\ref{Fig_SrMn2Sb2_MT}(a) shows the zero-field-cooled (ZFC) magnetic susceptibility $\chi(T)\equiv M(T)/H$ of \sms\ in a small magnetic field $H = 0.1$~T applied in the $ab$~plane ($\chi_{ab}$) and along the $c$~axis ($\chi_c$). These data exhibit a clear AFM transition at $T\rm_N = 110(5)$~K\@. The value of $T\rm_N$ is confirmed from a plot of $d\chi_{ab}(T)/dT$ versus $T$ in the inset of Fig.~\ref{Fig_SrMn2Sb2_MT}(a). The anisotropy of $\chi$ below $T\rm_N$ indicates that the $c$~axis is the hard axis and hence that the ordered moment lies in the $ab$~plane. The substantial value of $\chi_{ab}$ for $T\to0$ indicates that the AFM structure of \sms\ could be either a collinear AFM with multiple domains aligned within the $ab$~plane or an intrinsically noncollinear structure; this behavior is similar to that of the isostructural compound \sma\ \cite{Sangeetha2016} that was ultimately found to have a collinear AFM structure with the ordered moments aligned in the $ab$~plane, but with three approximately equally-populated domains at an angle of 60$^\circ$ to each other which led to $\chi_{ab}(T\to0) \approx \chi(T_{\rm N})/2$ \cite{Das2017}, as also observed in Fig.~\ref{Fig_SrMn2Sb2_MT}(a).

Figure~\ref{Fig_SrMn2Sb2_MT}(b) shows that the anisotropy in $\chi(T)$ almost disappears above~$T_{\rm N}$ in a field of 3~T which indicates that the  magnetocrystalline anisotropy in \sms\ is rather small. These results also show that contrary to a three-dimension AFM for which $\chi$ decreases above $T_{\rm N}$, the data for \sms\ increase above $T_{\rm N}$, reach a broad maximum at $\sim 300$~K and then slowly decrease.  A Curie-Weiss behavior is not attained up to 900~K, i.e., positive curvature in $\chi(T)$ is not clearly evident, indicating that strong dynamic AFM fluctuations occur up to at least 900~K\@.  Within a local-moment picture, these features at $T>T_{\rm N}$ are characteristic of a quasi-two-dimensional AFM\@.

$M(H)$ isotherms for a single crystal of \sms\ with $H\parallel ab$ and $H\parallel c$ are shown in Figs.~\ref{Fig:SrMn2Sb2_MH}(a) and~\ref{Fig:SrMn2Sb2_MH}(b), respectively.  The data for $H\parallel c$ are proportional to~$H$ at all temperatures, indicating lack of significant ferromagnetic or saturable paramagnetic impurities.  However, for $H\parallel ab$, the magnetization shows {\it positive} curvature for $H\lesssim 1$~T at low temperatures.  An expanded plot of the data with $H\leq 3$~T for $T = 100$~K and 1.8~K is shown in the inset of Fig.~\ref{Fig:SrMn2Sb2_MH}(a), where the positive curvature is somewhat more apparent.  One can greatly amplify the low-field $M(H)$ behaviors of \sms\ by plotting $M/H$ versus~$H$ as shown for $H\parallel ab$ and $H\parallel c$ in Figs.~\ref{Fig:SrMn2Sb2_MonH}(a) and~\ref{Fig:SrMn2Sb2_MonH}(b), respectively.  If $M_{ab}\propto H$ at low fields, the result would just be a horizontal line.  Instead, one sees in Fig.~\ref{Fig:SrMn2Sb2_MonH}(a) a strong increase in $M_{ab}/H$ with increasing~$H$ up to 1~T, where it levels out.  This indicates that a continuous $ab$-plane metamagnetic transition for $H\parallel ab$ with $0<H\lesssim 1$~T occurs at low temperatures $T\leq 50~{\rm K}<T_{\rm N}$ but not at $T \gtrsim 100~{\rm K} \approx T_{\rm N}$ or for $H\parallel c$ at all temperatures from 1.8 to 350~K\@.  Thus this transition is associated with the AFM state.  Determining the nature of this metamagnetic transition requires further investigation.

Our results disagree in two important ways with the $\chi(T)$ data for a polycrystalline sample of \sms\ in Fig.~2 of Ref.~\cite{Bobev2006}.  First, these authors reported a broad peak at $\sim 250$~K in $\chi(T)$ measured in 0.1~T which bears no resemblance to the powder average of the $\chi(T)$ data in $H=0.1$~T in our Fig.~\ref{Fig_SrMn2Sb2_MT}(a).  Second, the average of their data from $T=5$ to 300~K is $\approx 0.05$~cm$^3$/mole-\sms, whereas the value of the powder-averaged $\chi$ up to 300~K in our Fig.~\ref{Fig_SrMn2Sb2_MT}(a) is $\approx 4\times10^{-3}$~cm$^3$/mole-\sms, more than an order of magnitude smaller.  We infer that their results were strongly affected by FM MnSb impurities which have a strongly composition-dependent Curie temperature $T_{\rm C}$ that can vary between 100 and 320~K \cite{Okita1968, Chen1977} and with a large ordered moment of $\sim 3~\mu_{\rm B}$/Mn \cite{Okita1968}.  We prevented the formation of MnSb impurities when growing our \sms\ crystals by using excess Sr in the crystal growth, as noted in Sec.~\ref{ExpDetails}.

The presence of FM MnAs impurities with $T_{\rm C} = 317$~K in the early ${\rm BaMn_2As_2}$ crystals made it tedious to obtain the intrinsic anisotropic $\chi(T)$ of this compound because the contribution of these impurities to the measured magnetization had to be accounted for by taking $\chi$ to be the high-field slope of an $M(H)$ isotherm that was measured at each temperature for each field direction~\cite{Singh2009}.  The small amount of MnAs impurities present in those crystals did not contribute significantly to the magnetization when the measurement temperature was above $T_{\rm C}$ of MnAs.

\subsubsection{\bms}

\begin{figure}
\includegraphics[width=3.3in]{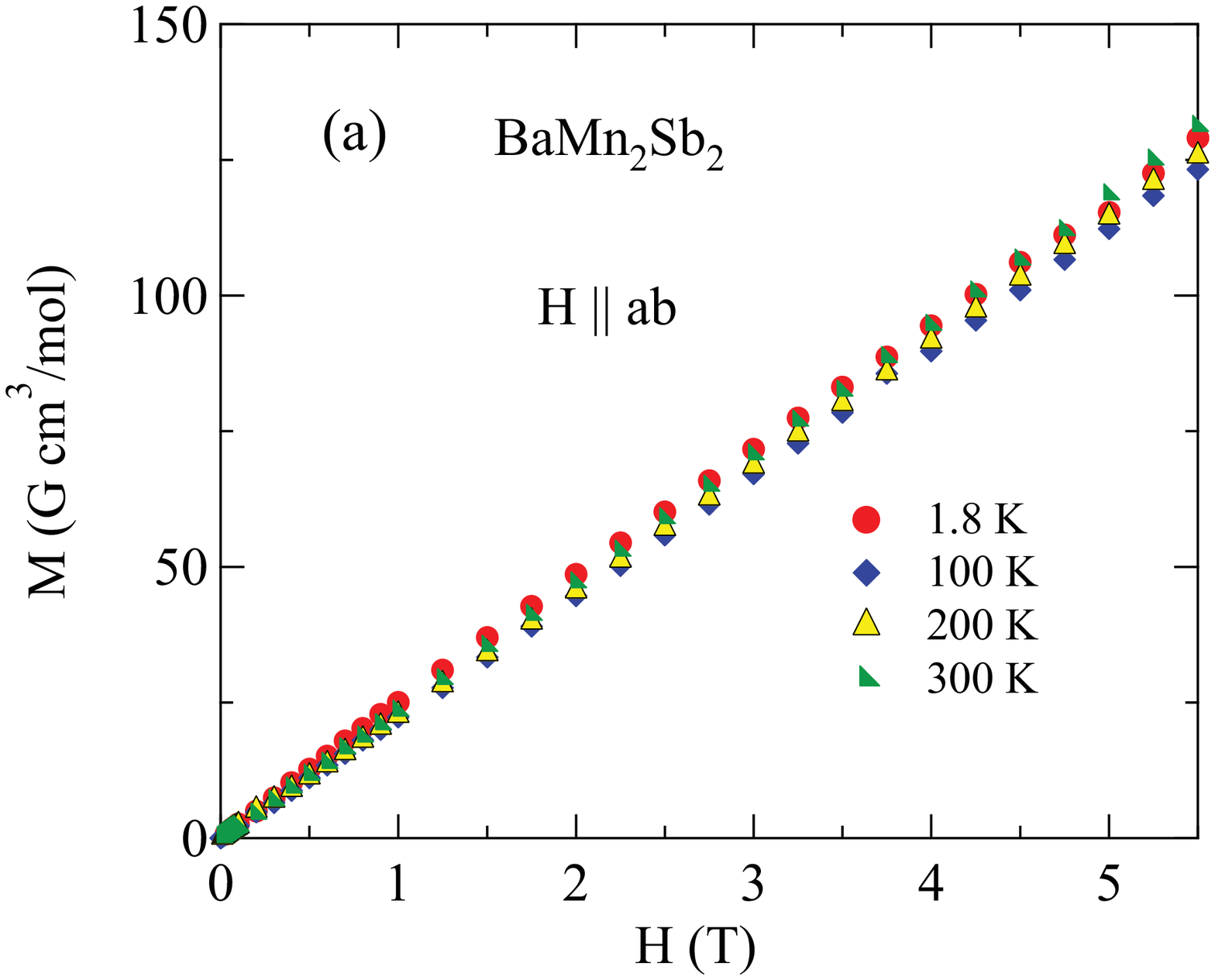}
\includegraphics[width=3.3in]{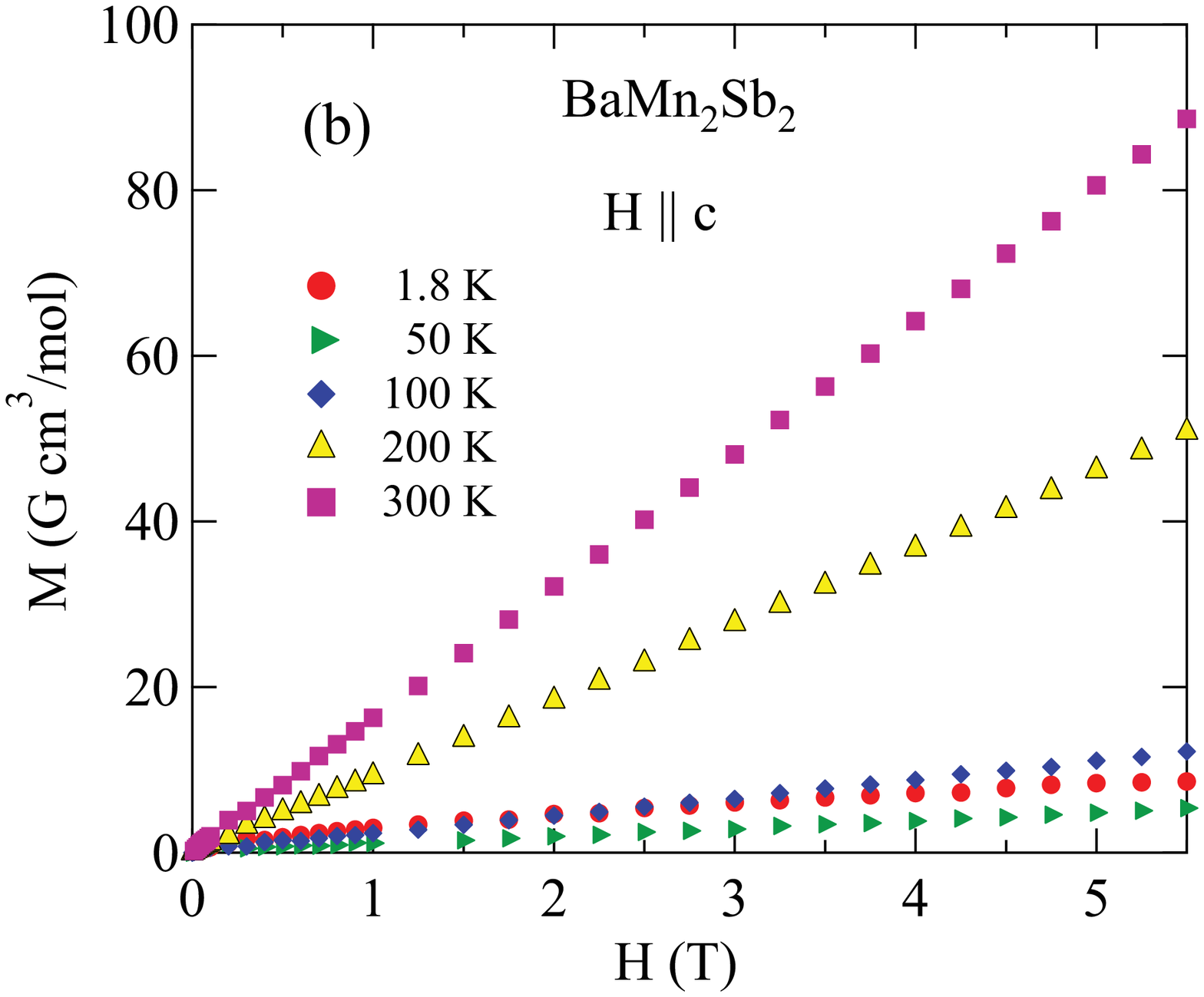}
\caption{(Color online) Isothermal magnetization $M$ of \bms\ as a function of magnetic field $H$ at the temperatures listed with (a)~$H$ in the $ab$~plane ($H\,||\, ab$) and (b)~$H$ along $c$~axis ($H\,||\, c$). Note the different scales on the ordinates in (a) and~(b).}
\label{Fig:BaMn2Sb2_MH} 
\end{figure}

\begin{figure}
\includegraphics[width=3.3in]{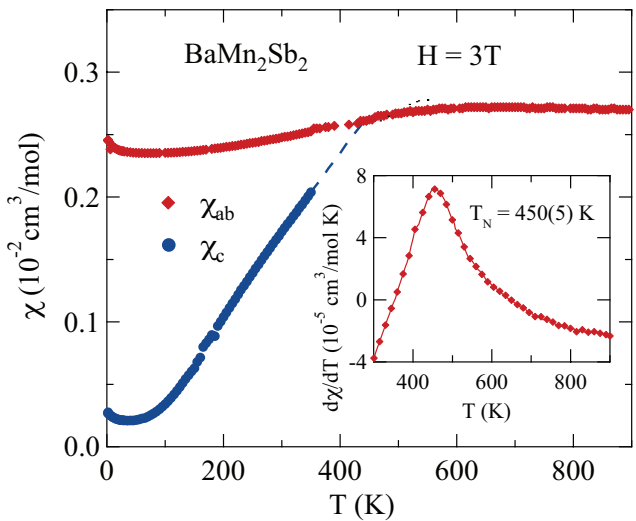}
\caption{(Color online) ZFC magnetic susceptibility $\chi(T)$ of \bms\ as a function of temperature $T$ for 1.8 to 350~K in $H=3$~T applied along the $c$ axis and $\chi(T)$ for 1.8$\leq T \leq$ 900~K in $H=3$~T applied in the $ab$-plane. The inset shows the derivative $d\chi_{ab}/dT$ versus~$T$ used to estimate $T_{\rm N}$ as the temperature of the peak.}
\label{Fig_BaMn2Sb2_MT}
\end{figure}

Plots of the the isothermal magnetization $M$ of a \bms\ single crystal versus~$H$ for $H\,||\,ab$ and $H\,||\,c$ at various temperatures are shown in Figs.~\ref{Fig:BaMn2Sb2_MH}(a) and \ref{Fig:BaMn2Sb2_MH}(b), respectively. Anisotoropic behavior between $H \parallel ab$ and $H \parallel c$ is evident at low temperatures.  One sees from these plots that the contribution of ferromagnetic or saturable paramagnetic impurities to the magnetization data is insignificant at all measured temperatures.

The zero-field-cooled (ZFC) magnetic susceptibilities $\chi\equiv M/H$ of \bms\ versus $T$ from 1.8 to 300~K with $H=3$~T applied along the $c$~axis ($H \parallel c,\ \chi_c$), and $\chi(T)$ for $T$ between 1.8 to 900~K with $H=3$~T applied in the $ab$~plane ($H \parallel ab,\ \chi_{ab}$), are shown in Fig.~\ref{Fig_BaMn2Sb2_MT}. The anisotropy between the $c$~axis and $ab$~plane data below $\sim400$~K is a textbook example of collinear long-range AFM ordering with the ordered moments oriented along the $c$~axis \cite{Johnston2012, Johnston2015}.  Linearly extrapolating the $\chi_c(T)$ data to higher temperatures to intersect the $\chi_{ab}(T)$ data (dashed line) suggests $T_{\rm N}\sim430$~K\@.  A more accurate estimate is obtained from a plot of the derivative $d\chi_{ab}/dT$ versus~$T$ in the inset of Fig.~\ref{Fig_BaMn2Sb2_MT}.  A clear peak in these data gives the N\'eel temperature as $T_{\rm N} = 450(5)$~K\@.

Figure~\ref{Fig_BaMn2Sb2_MT} further shows that $\chi_{ab}(T)$ does not decrease above $T_{\rm N}$ as would be expected for a three-dimensional AFM, but instead increases and appears to approach a maximum at $\sim 600$~K, followed by a slow decrease.  This paramagnetic behavior is similar to that of \sms\ discussed above and is characteristic of strong short-range dynamic two-dimensional AFM fluctuations above $T_{\rm N}$ \cite{Johnston2011} as seen previously in, e.g., tetragonal \bma\ \cite{Johnston2011} and trigonal \cma\ and \sma\ \cite{Sangeetha2016}.  

Our overall $\chi(T)$ and $M(H)$ results for \bms\ are similar to those for other isostructural Mn compounds such as \bma\ \cite{Singh2009}, BaMn$_2$P$_2$ \cite{Brock1994}, (Ca,Ba)Mn$_2$Ge$_2$ \cite{Malaman1994}, and BaMn$_2$Bi$_2$ \cite{Saparov2013}.  Electronic structure studies of a G-type collinear AFM structure in \bms\ were carried out previously and the ordered Mn moment at $T=0$ was calculated to be $\approx 3.8~\mu_{\rm B}$/Mn \cite{{Xia2008}} or $3.55~\mu_{\rm B}$/Mn \cite{An2009}.

\subsection{\label{Sec:HC}  Heat Capacity}

\begin{figure}
\includegraphics[width=3.3in]{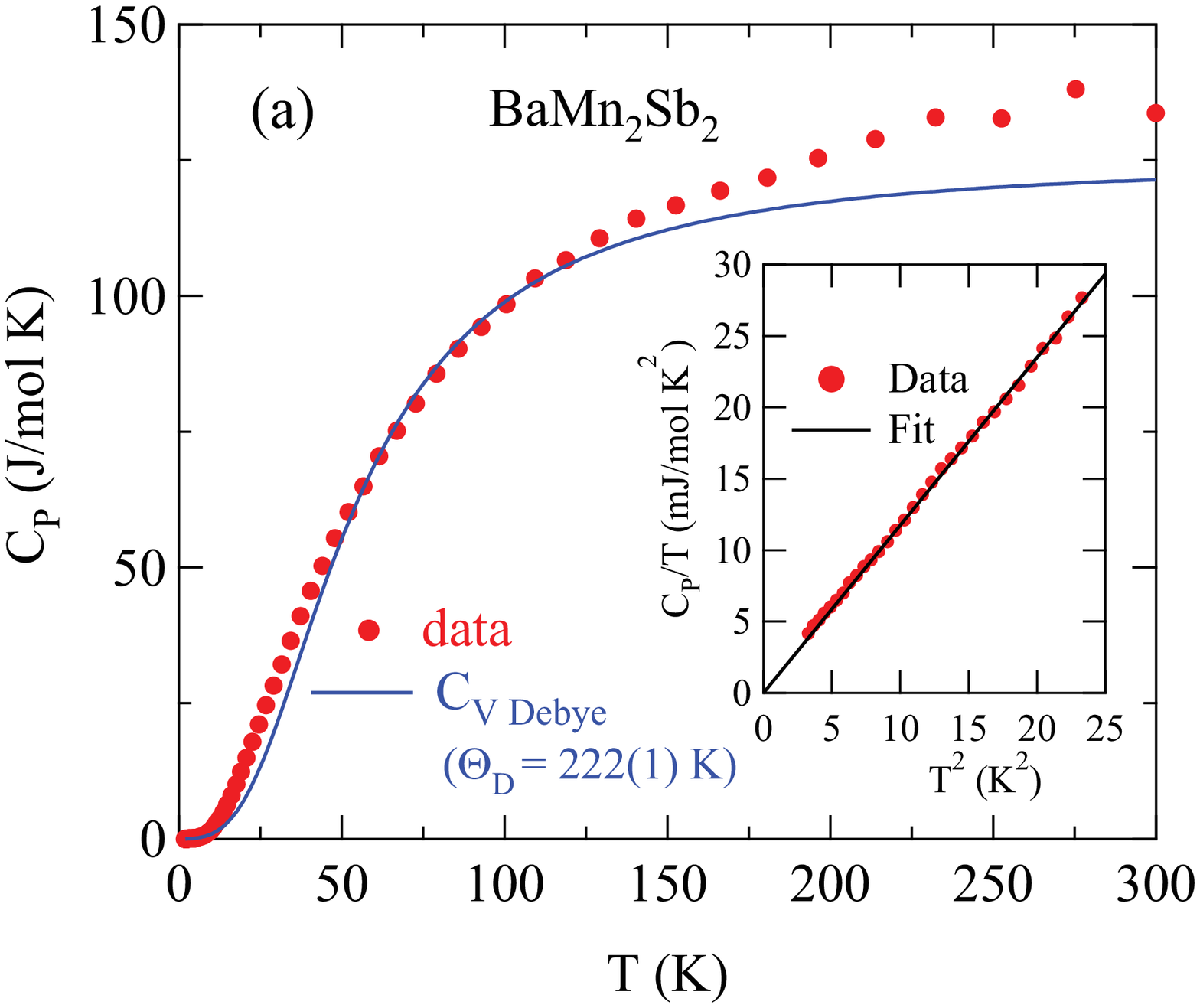}
\includegraphics[width=3.3in]{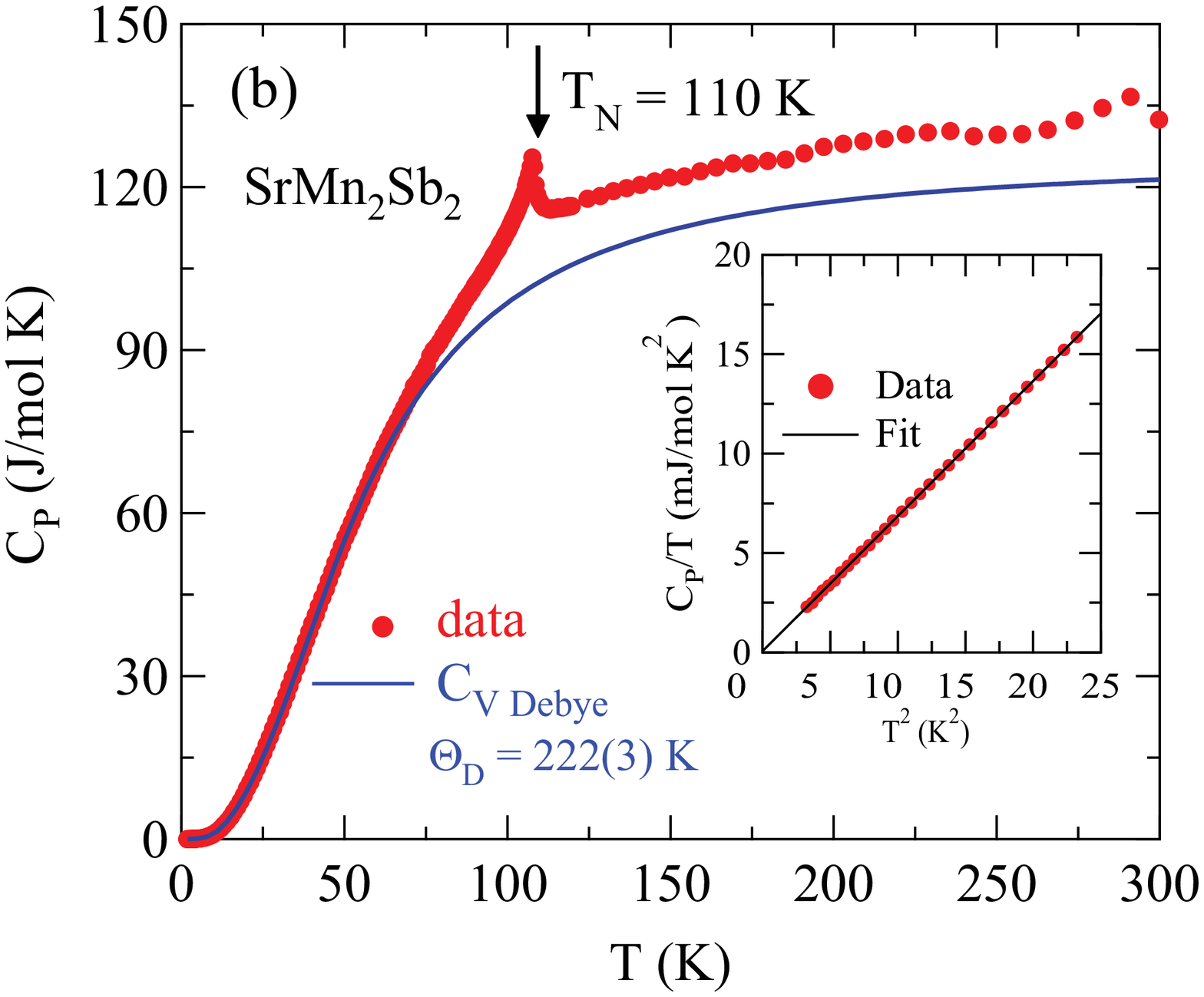}
\caption{(Color online) Heat capacity $C{\rm_p}$ versus temperature $T$ and fits of $C_{\rm V\,Debye}(T)$ in Eqs.~(\ref{Eq:Debye_Fit}) to the data over restricted temperature intervals (see text) for (a) \bms\ and (b) \sms\@. The fits and extrapolations are shown as blue curves, and the fitted Debye temperatures $\Theta_{\rm D}$ are shown in each panel.  The respective insets show $C{\rm_p}(T)/T$ versus $T^2$ for $T< 5$~K, where the straight lines though the respective data are fits by Eq.~(\ref{Eq:ConT}) where the fitted parameters are given in Eqs.~(\ref{Eqs:bmsFitPars}) and~(\ref{Eqs:smsFitPars}), respectively.}
\label{Fig:BaSrMn2Sb2_HC}
\end{figure}

The $C{\rm{_p}}(T)$ data for \bms\ and \sms\ in the temperature range between 1.8 and 300~K are shown in Figs.~\ref{Fig:BaSrMn2Sb2_HC}(a) and~\ref{Fig:BaSrMn2Sb2_HC}(b), respectively. The temperature of the sharp $\lambda$-type anomaly in  $C{\rm{_p}}(T)$ for \sms\ at 110~K is in good agreement with the $T\rm_N$ found from the above $\chi(T)$ data for this material. The insets in Figs.~\ref{Fig:BaSrMn2Sb2_HC}(a) and~~\ref{Fig:BaSrMn2Sb2_HC}(b) show a proportional $C{\rm{_p}}/T$ versus $T^2$ behavior for 1.8 $\leq T \leq$ 5~K\@.  Hence we fit these data by the expression
\be
\frac{C{\rm_p}}{T}=\gamma+\beta T^{2},
\label{Eq:ConT}
\ee
where the coefficient $\gamma$ in metals is due to the conduction electron contribution (Sommerfeld coefficient) if nonzero and $\beta$ reflects the low-$T$ Debye $T^3$ lattice contribution together with a three-dimensional AFM spin-wave contribution if present. 
For \bms, the fit yields
\bse
\label{Eqs:bmsFitPars}
\bea
 \gamma &=& 0.01(1)~\rm mJ/mol~K^2,\\*
  \beta &=& 1.17(2)~\rm mJ/mol~K^4,
\eea
\ese
and for \sms\ we obtain
\bse
\label{Eqs:smsFitPars}
\bea
 \gamma &=& 0.02(4)~\rm mJ/mol~K^2, \\
 \beta  &=& 0.680(3)~\rm mJ/mol~K^4.
\eea
\ese
The null values of $\gamma$ are consistent with the insulating ground states found from the respective $\rho(T)$ measurements in Sec~\ref{Sec:Rho}.   Assuming no contribution from 3D AFM spin waves, we estimate the Debye temperatures ($\Theta{\rm_D}$) of the two compounds from the $\beta$ values using the expression
\be
\Theta{\rm_D}=\left(\frac{12\pi^{4}Rn}{5\beta}\right)^{1/3}
\label{Eq_thetaD}
\ee
where $R$ is the molar gas constant and $n$ is the number of atoms per formula unit ($n=5$ for \bms\ and \sms). We thus obtain 
\bse
\bea
\Theta{\rm_D} &=& 202(1)~\rm K \quad   ~~(BaMn_2Sb_2),\label{Eq:QC_BMS}\\
 &=&242.5(3)~\rm{K}\quad(SrMn_2Sb_2).\label{Eq:SMSCpFit}
\eea
\ese

The lattice contributions $C_{\rm latt}(T)$ to $C_{\rm p}(T)$ for \bms\ and \sms\ at higher temperatures were obtained by fitting the respective $C_{\rm p}(T)$ data over specified temperature ranges where the magnetic contribution $C_{\rm mag}(T)$ was expected to be small by the molar heat capacity expression 
\bse
\label{Eq:CplatticeFit}
\be
C{\rm_{latt}} = n~C_{\rm {V\,Debye}} 
\ee
where $n$ is defined above.  Here $C\rm_{V\,Debye}$ is the Debye lattice heat capacity per mole of atoms given by
\be
C{\rm_{V\,Debye}}(T) = 9R \left(\frac{T}{\Theta_D}\right)^3\int_{0}^{\Theta_D/T}\frac{x^4e^x}{(e^x-1)^2} dx
\label{Eq:Debye_Fit}
\ee
\ese
where $\Theta\rm_D$ is the Debye temperature. An accurate analytic Pad\'e approximant was used for the Debye function in Eq.~(\ref{Eq:Debye_Fit}) for fitting the $C{\rm_p}(T)$ data \cite{Goetsch2012}.  The fit of $C{\rm_p}(T)$ for \bms\ by Eqs.~(\ref{Eq:CplatticeFit}) over the temperature range from 50 to 100~K and its extrapolation to higher temperatures is shown by the solid blue curve in Fig.~\ref{Fig:BaSrMn2Sb2_HC}(a), yielding $\Theta\rm_D$ = 222(1)~K. This value of $\Theta\rm_D$ is comparable with the value of 202~K in Eq.~(\ref{Eq:QC_BMS}) determined from the fit of the $C_{\rm_p}$ for \bms\ at low $T$ by Eq.~(\ref{Eq:ConT}).

For \sms,\ the $C_{\rm_p}(T)$ data for $1.8\leq T \leq10$~K and $50\leq T \leq 70$~K were fitted simultaneously by Eqs.~(\ref{Eq:CplatticeFit}).  The fit and its interpolation and extrapolation are shown by the blue curve in Fig.~\ref{Fig:BaSrMn2Sb2_HC}(b) obtained using the fitted Debye temperature $\Theta\rm_D$ = 222(3)~K\@. Again, the value of $\Theta\rm_D$ for \sms\ is comparable with the value of 242.5~K in Eq.~(\ref{Eq:SMSCpFit}) determined from the fit of the $C_{\rm_p}(T)$ data at low $T$ by Eq.~(\ref{Eq:ConT})

\begin{figure}
\includegraphics[width=3.4in]{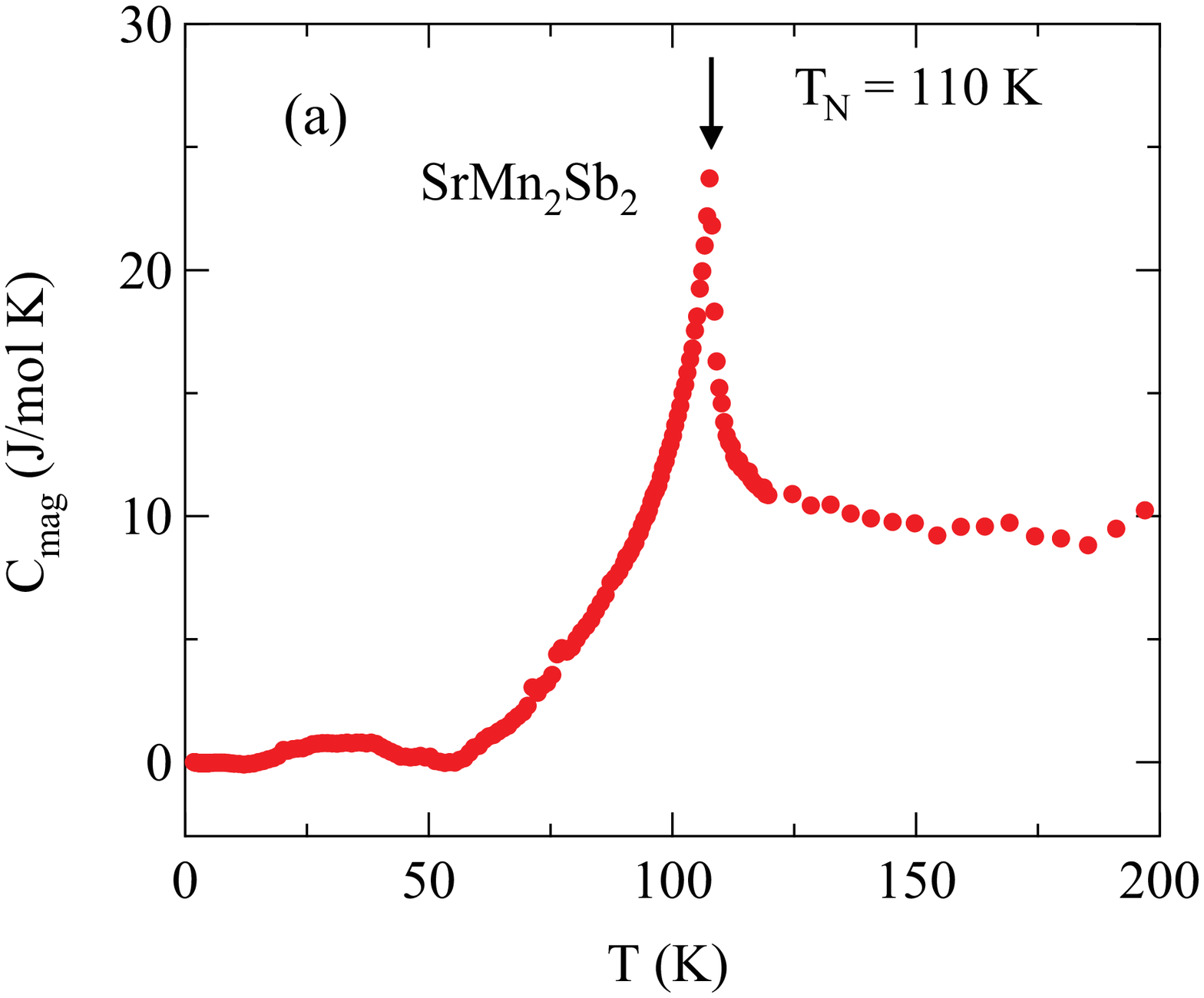}\vspace{0.2in}
\includegraphics[width=3.4in]{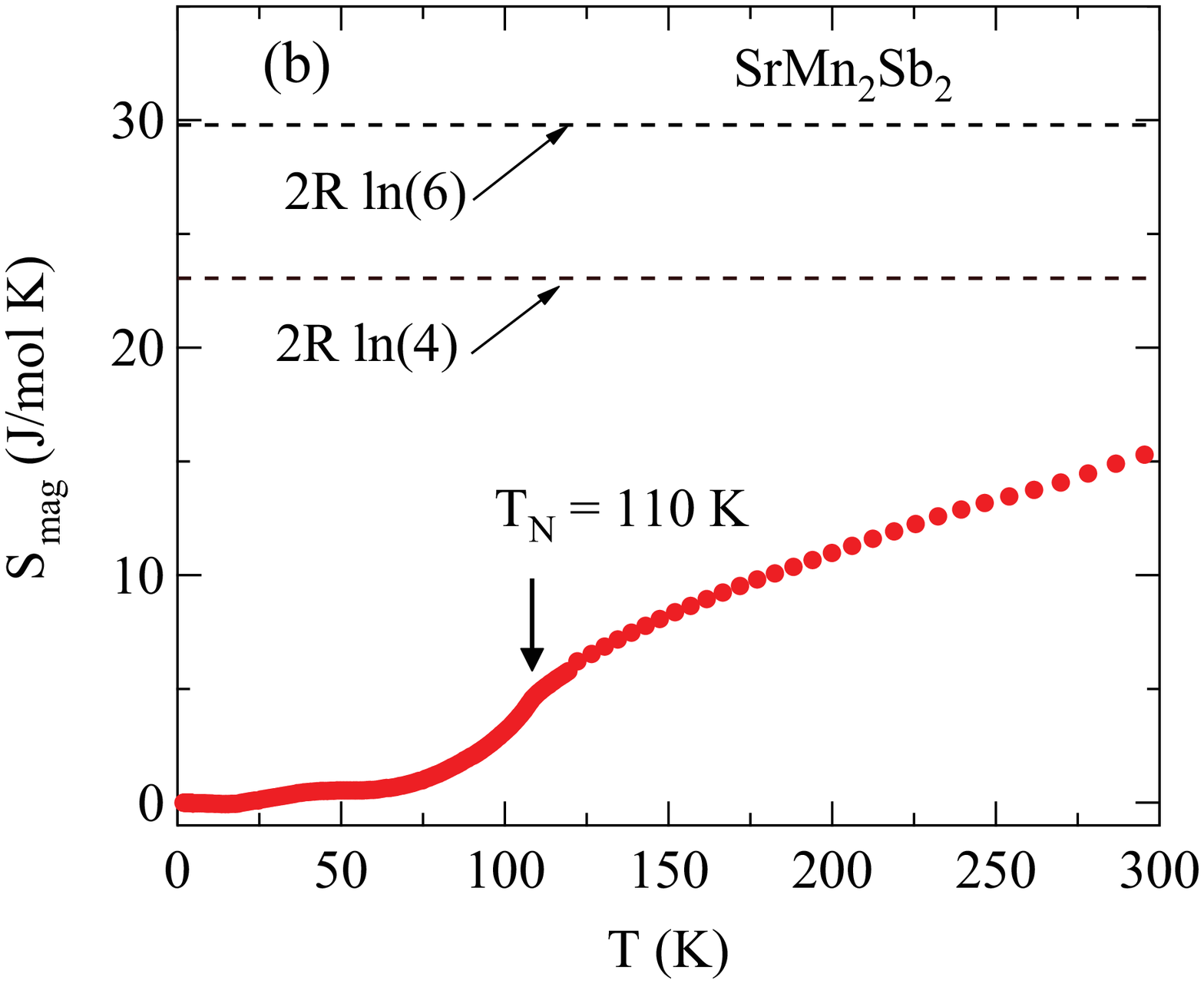}
\caption{(Color online) (a)~Molar magnetic heat capacity $C_{\rm mag}$ versus temperature~$T$ for \sms\ obtained using Eq.~(\ref{Eq:Cmag}). The $\lambda$ anomaly at $T_{\rm N} = 110$~K arises from AFM order with the moments aligned in the $ab$~plane.  (b)~Molar magnetic entropy $S_{\rm mag}(T)$ of \sms\ obtained using the data in~(a) and Eqs.~(\ref{Eq:Smag}). The dashed lines are $S_{\rm mag}(T\to\infty)$ for $S=5/2$ and $S=3/2$ in Eqs.~(\ref{Eqs:EntropyValues}), as indicated. }
\label{Fig:SrMn2Sb2_CSmag}
\end{figure}

The fitted $C{\rm_{latt}}(T)$ in Fig.~\ref{Fig:BaSrMn2Sb2_HC}(b) is used to obtain an estimate of $C{\rm_{mag}}(T)$ to the measured heat capacity of \sms\@. The molar magnetic heat capacity is then obtained from
\be
C{\rm_{mag}}(T) = C{\rm{_p}}(T) - C{\rm_{latt}}(T).
\label{Eq:Cmag}
\ee
The $C{\rm_{mag}}(T)$ for \sms\ obtained using Eq.~(\ref{Eq:Cmag}) is shown in Fig.~\ref{Fig:SrMn2Sb2_CSmag}(a), where a sharp peak at $T\rm_N$ = 110~K is seen. The small bump below $T\rm_N$ is an artifact and the nonzero $C\rm_{mag}$ above $T\rm_N$ indicates strong dynamic short-range AFM order above $T_{\rm N}$\@. 

The magnetic entropy $S{\rm_{mag}}(T)$ for \sms\ is calculated from $C{\rm_{mag}}(T)$ in Fig.~\ref{Fig:SrMn2Sb2_CSmag}(a)  using
\be
S{\rm_{mag}}(T) =\int_{0}^{T} \frac{C{\rm_{mag}}(T)}{T} dT,
\label{Eq:Smag}
\ee
and the result is shown in Fig.~\ref{Fig:SrMn2Sb2_CSmag}(b). The entropy of completely disordered spins $S$ per mole of \sms\ is $S(T\to\infty) = 2 R \ln(2S+1)$, which gives
\bse
\label{Eqs:EntropyValues}
\label{Eqs:SmagVals}
\bea
S_{\rm mag}(T\to\infty) &=& 23.1 \,{\rm \frac{J}{mol\,K}}\quad (S=3/2)\\*
&=& 29.8\,{\rm \frac{J}{mol\,K}}\quad (S=5/2),
\eea
\ese
as shown by the horizontal dashed black lines in Fig.~\ref{Fig:SrMn2Sb2_CSmag}(b).  The $S_{\rm mag}(300~{\rm K}) \approx 15~{\rm J/mol\,K}$ for \sms\ in Fig.~\ref{Fig:SrMn2Sb2_CSmag}(b) is only $\approx 50$\% of the value for $S=5/2$ in Eqs.~(\ref{Eqs:SmagVals}) and is still only $\approx 65$\% of the value for $S=3/2$. The remaining entropy is evidently not released until much higher temperatures.  Thus the strong dynamic short-range AFM order revealed in the $C_{\rm mag}(T)$ and $S_{\rm mag}(T)$ data above $T_{\rm N}$ is consistent with the above conclusion from the $\chi(T)$ data that the dynamic short-range AFM order survives from $T_{\rm N}$ up to at least 900~K\@.

\section{\label{Sec:Summary} Summary}

Refinements of our x-ray diffraction data confirm that \sms\ crystallizes in the trigonal \cas-type structure with space group $P\bar{3}m1$, whereas \bms\ adopts the body-centered tetragonal \tcs\ structure with space group $I4/mmm$.  Heat capacity and electrical resisitivity $\rho(T)$ measurements indicate that both compounds are insulators at low temperatures.  At higher temperatures, the $\rho(T)$ data show that \sms\ and \bms\ are semiconductors at with activation energies estimated to be $\gtrsim0.35$ and 0.16~eV, respectively.

Long-range antiferromagnetic (AFM) order arising from the nominal Mn$^{+2}$ spins with $d^5$ electronic configuration and spin $S=5/2$ is reported for single crystals of \sms\ and \bms\ at $T\rm_N$ = 110~K and 450~K, respectively, as determined from magnetic susceptibility $\chi(T)$ and heat capacity $C_{\rm p}(T)$ measurements.  We infer from the anisotropy in the $\chi(T)$ measurements below $T_{\rm N}$ that the ordered moments in \sms\ lie in the $ab$~plane and exhibit either intrinsic planar noncollinear AFM order or extrinsic noncollinear AFM order arising from multiple collinear AFM $ab$-plane domains.  From magnetization versus field isotherms, we discovered that \sms\ exhibits a continuous metamagnetic transition at low fields between 0 and 1~T\@.  The nature of this transition remains to be determined.  On the other hand, the classic anisotropy of the $\chi(T)$ data for \bms\ at temperature below $T_{\rm N}$ indicates that this compound exhibits collinear AFM order with the moments aligned along the $c$~axis.

The $\chi(T)$ above $T\rm_N$ in \sms\ is isotropic and for both \sms\ and \bms, indicating weak magnetocrystalline anisotropy.  Both compounds exhibit broad maxima in $\chi(T)$ above $T\rm_N$ as shown in Figs.~\ref{Fig_SrMn2Sb2_MT}(b) and~\ref{Fig_BaMn2Sb2_MT}, respectively, and do not settle into a Curie-Weiss behavior up to 900~K\@. Furthermore, the magnetic entropy of \sms\ obtained at 300~K from the heat capacity measurements is only half that expected for disordered Mn spins-5/2.  These results indicate that within a local-moment picture, \sms\ and \bms\ exhibit strong dynamic AFM fluctuations up to at least 900~K and hence are quasi-two-dimensional (2D) local-moment AFMs \cite{Johnston2011}.

The differences between the smaller intraplanar and larger interplanar interatomic distances between Mn atoms in Table~\ref{Table_SXRD} and the resultant differences between their interactions in \sms\ and \bms\ are expected to strongly influence their magnetic properties. In both compounds, the smallest intraplanar distances between the Mn atoms are much smaller than the smallest interlayer distances. This apparently leads to the 2D dynamic short-range correlations in \sms\ and \bms\ above $T\rm_N$.  The differences in the structures of the Mn layers in \sms\ and \bms\ also appear to be important.  From Table~\ref{Table:MnPnMagProps}, these differences may contribute to the large differences in N\'eel temperatures of layered Mn pnictides between the \cas\ and \tcs\ crystal structures.  Theoretical studies of the origins of these behaviors would be most interesting.

\acknowledgments

This research was supported by the U.S. Department of Energy, Office of Basic Energy Sciences, Division of Materials Sciences and Engineering.  Ames Laboratory is operated for the U.S. Department of Energy by Iowa State University under Contract No.~DE-AC02-07CH11358.


\clearpage

\begin{thebibliography}{99}

\bibitem{Kamihara} Y. Kamihara, T. Watanabe, M. Hirano, and H. Hosono, Iron-Based Layered Superconductors La(O$_{1-x}$F$_x$)FeAs ($x=0.05$--0.12) with $T\rm_c=$ 26~K, J. Am. Chem. Soc. {\bf 130}, 3296 (2008).

\bibitem{Just1996} G. Just and P. Paufler, On the coordination of \tcs\ (BaAl$_4$)-type compounds within the field of free parameters, J. Alloys Compd. {\bf 232}, 1 (1996).

\bibitem{Johnston2010} D. C. Johnston, The puzzle of high temperature superconductivity in layered iron pnictides and chalcogenides, Adv. Phys. {\bf 59}, 803 (2010).

\bibitem{Stewart2011} G. R. Stewart, Superconductivity in iron compounds, Rev. Mod. Phys. {\bf 83}, 1589 (2011).

\bibitem{Scalapino2012} D. J. Scalapino, A common thread: The pairing interaction for unconventional superconductors, Rev. Mod. Phys. {\bf 84}, 1383 (2012).

\bibitem{Wang2008} C. Wang. L. Li, S. Chi, Z. Zhu, Z. Ren, Y. Li, Y. Wang, X. Lin, Y. Luo, S. Jiang, X. Xu, G. Cao, and Z. Xu, Thorium-doping-induced superconductivity up to 56~K in Gd$_{1-x}$Th$_x$FeAsO, Europhys. Lett. {\bf 83}, 67006 (2008).

\bibitem{Sefat2009} A. S. Sefat, D. J. Singh, R. Jin, M. A. McGuire, B. C. Sales, and D. Mandrus, Renormalized behavior and proximity of BaCo$_2$As$_2$ to a magnetic quantum critical point, Phys. Rev. B {\bf 79}, 024512 (2009).

\bibitem{Ohta2009} H. Ohta and K. Yoshimura, Anomalous magnetization in the layered itinerant ferromagnet LaCoAsO, Phys. Rev. B {\bf 79}, 184407 (2009).

\bibitem{Cheng2012} B. Cheng, B. F. Hu, R. H. Yuan, T. Dong, A. F. Fang, Z. G. Chen, G. Xu, Y. G. Shi, P. Zheng, J. L. Luo, and N. L. Wang, Field-induced spin-flop transitions in single-crystalline CaCo$_2$As$_2$, Phys. Rev. B {\bf 85}, 144426 (2012).

\bibitem{Quirinale2013} D. G. Quirinale, V. K. Anand, M. G. Kim, A. Pandey, A. Huq, P. W. Stephens, T. W. Heitmann, A. Kreyssig, R. J. McQueeney, D. C. Johnston, and A. I. Goldman, Crystal and magnetic structure of CaCo$_{1.86}$As$_2$ studied by x-ray and neutron diffraction, Phys. Rev. B {\bf 88}, 174420 (2013).

\bibitem{Anand2014_CCA} V. K. Anand, R. S. Dhaka, Y. Lee, B. N. Harmon, A. Kaminski, and D. C. Johnston, Physical properties of metallic antiferromagnetic CaCo$_{1.86}$As$_2$, Phys. Rev. B {\bf 89}, 214409 (2014).

\bibitem{Anand2014_BCA} V. K. Anand, D. G. Quirinale, Y. Lee, B. N. Harmon, Y. Furukawa, V. V. Ogloblichev, A. Huq, D. L. Abernathy, P. W. Stephens, R. J. McQueeney, A. Kreyssig, A. I. Goldman, and D. C. Johnston, Crystallography and physical properties of BaCo$_2$As$_2$, Ba$_{0.94}$K$_{0.06}$Co$_2$As$_2$ and Ba$_{0.78}$K$_{0.22}$Co$_2$As$_2$, Phys. Rev. B {\bf 90}, 064517  (2014).

\bibitem{Pandey2013}  A. Pandey, D. G. Quirinale, W. Jayasekara, A. Sapkota, M. G. Kim, R. S. Dhaka, Y. Lee, T. W. Heitmann, P. W. Stephens, V. Ogloblichev, A. Kreyssig, R. J. McQueeney, A. I. Goldman, A. Kaminski, B. N. Harmon, Y. Furukawa, and D. C. Johnston, Crystallographic, electronic, thermal, and magnetic properties of single-crystal SrCo$_2$As$_2$, Phys. Rev. B {\bf 88}, 014526 (2013).

\bibitem{Jayasekara2013} W. Jayasekara, Y. Lee, A. Pandey, G. S. Tucker, A. Sapkota, J. Lamsal, S. Calder, D. L. Abernathy, J. L. Niedziela, B. N. Harmon, A. Kreyssig, D. Vaknin, D. C. Johnston, A. I. Goldman, and R. J. McQueeney, Stripe Antiferromagnetic Spin Fluctuations in SrCo$_2$As$_2$, Phys. Rev. Lett. {\bf 111}, 157001 (2013).

\bibitem{Sangeetha2016b} N. S. Sangeetha, E. Cuervo-Reyes, A. Pandey, and D. C. Johnston, EuCo$_2$P$_2$: A model molecular-field helical Heisenberg antiferromagnet, Phys. Rev. B {\bf 94}, 014422 (2016).

\bibitem{Park2013} S.-W. Park, H. Mizoguchi, K. Kodama, S.-I. Shamoto, T. Otomo, S. Matsuishi, T. Kamiya, and H. Hosono, Magnetic Structure and Electromagnetic Properties of LnCrAsO with a ZrCuSiAs-type Structure (Ln = La, Ce, Pr, and Nd), Inorg. Chem. {\bf 52}, 13363 (2013).

\bibitem{DJSingh} D. J. Singh, A. S. Sefat, M. A. McGuire, B. C. Sales, D. Mandrus, L. H. VanBebber, and V. Keppens. Itinerant antiferromagnetism in BaCr$_2$As$_2$: Experimental characterization and electronic structure calculations, Phys. Rev. B {\bf 79}, 094429 (2009).

\bibitem{Filsinger} K. A. Filsinger, W. Schnelle, P. Adler, G. H. Fecher, M. Reehuis, A. Hoser, J.-U. Ho↵mann, P. Werner, M. Greenblatt, and C. Felser, Antiferromagnetic structure and electronic properties of BaCr$_2$As$_2$ and BaCrFeAs$_2$, Phys. Rev. B {\bf 95}, 184414 (2017).

\bibitem{Naumov2017}  P. G. Naumov, K. Filsinger, O. I. Barkalov, G. H. Fecher, S. A. Medvedev, and C. Felser, Pressure-induced transition to the collapsed tetragonal phase in ${\rm BaCr_2As_2}$, Phys. Rev. B {\bf 95}, 144106 (2017).

\bibitem{Richard} P. Richard, A. van Roekeghem, B. Q. Lv, T. Qian, T.K. Kim, M. Hoesch, J.-P. Hu, A. S. Sefat, S. Biermann, and H. Ding, Is BaCr$_2$As$_2$ symmetrical to BaFe$_2$As$_2$ with respect to half 3$d$ shell filling?,  Phys. Rev. B {\bf 95}, 184516 (2017).

\bibitem{Paramanik} U. B. Paramanik, R. Prasad, C. Geibel, and Z. Hossain, Itinerant and local-moment magnetism in EuCr$_2$As$_2$ single crystals, Phys. Rev. B. {\bf 89}, 144423 (2014).

\bibitem{Pfisterer1980} M. Pfisterer and G. Nagorsen, On the Structure of Ternary Arsenides, Z. Naturforsch. {\bf 35b}, 703 (1980).

\bibitem{Pfisterer1983} M. Pfisterer and G. Nagorsen, Bonding and Magnetic Properties in Ternary Arsenides ET$_2$As$_2$, Z. Naturforsch. {\bf 38b}, 811 (1983).

\bibitem{Das2017b} P. Das, N. S. Sangeetha, G. R. Lindemann, T. W. Heitmann, A. Kreyssig, A. I. Goldman, R. J. McQueeney, D. C. Johnston, and D. Vaknin, Itinerant G-type antiferromagnetic order in SrCr$_2$As$_2$, Phys. Rev. B {\bf 96}, 014411 (2017).

\bibitem{Brock1994} S. L. Brock, J. E. Greedan, and S. M. Kauzlarich, Resistivity and Magnetism of $A$Mn$_2$P$_2$ ($A$ = Sr, Ba): The Effect of Structure Type on Physical Properties, J. Solid State Chem. {\bf 113}, 303 (1994).

\bibitem{Singh2009} Y. Singh, A. Ellern, and D. C. Johnston, Magnetic, transport and thermal properties of single crystals of the layered arsenide \bma, Phys. Rev. B {\bf 79}, 094519 (2009).

\bibitem{Wang2011} Z. W. Wang, H. X. Yang, H. F. Tian, H. L. Shi, J. B. Lu, Y. B. Qin, Z. Wang, and J. Q. Li, Structural and physical properties of \sma, J. Phys. Chem. Solids \textbf{72}, 457 (2011).

\bibitem{Wang2009} H. F. Wang, K. F. Cai, L. Wang, C. W. Zhou, Synthesis and thermoelectric properties of \bms\ single crystals, J. Alloys Compd. {\bf 477}, 519 (2009).

\bibitem{Saparov2013} B. Saparov and A. S. Sefat, Crystals, magnetic and electronic properties of a new \tcs-type BaMn$_2$Bi$_2$ and K-doped compositions, J. Solid State Chem. {\bf 204}, 32 (2013).

\bibitem{Sangeetha2016} N. S. Sangeetha, A. Pandey, Z. A. Benson and D. C. Johnston, Strong magnetic correlations to 900~K in single crystals of the trigonal antiferromagnetic insulators \sma\ and \cma, Phys. Rev. B {\bf 94}, 094417 (2016).

\bibitem{Yanagi2009} H. Yanagi, T. Watanabe, K. Kodama, S. Iikubo, S. Shamoto, T. Kamiya, M. Hirano, and H. Hosono, Antiferromagnetic bipolar semiconductor LaMnPO with ZrCuSiAs-type structure, J. Appl. Phys. {\bf 105}, 093916 (2009).

\bibitem{McGuire2016}M. A. McGuire and V. O. Garlea, Short- and long-range magnetic order in LaMnAsO, Phys. Rev. B {\bf 93}, 054404 (2016).

\bibitem{QZhang2016} Q. Zhang, C. M. N. Kumar, W. Tian, K. W. Dennis, A. I. Goldman, and D. Vaknin, Structure and magnetic properties of $Ln$MnSbO ($Ln=$ La, Ce), Phys. Rev. B {\bf 93}, 094413 (2016).

\bibitem{Li2014} Y. Li, Y. Luo, L. Li, B. Chen, X. Xu, J. Dai, X. Yang, L. Zhang, G. Cao, and Z. Xu, Kramers non-magnetic superconductivity in LnNiAsO superconductors, J. Phys.: Condens. Matter {\bf 26}, 425701 (2014). 

\bibitem{Tegel2008}M. Tegel, D. Bichler, and D. Johrendt, Synthesis, crystal structure and superconductivity of LaNiPO, Solid State Sciences {\bf 10}, 193 (2008). 

\bibitem{Ronning2009}F. Ronning, N. Kurita, E. D. Bauer, B. L. Scott, T. Park, T. Klimczuk, R. Movshovich, and J. D. Thompson, Ni$\rm _2$X$_2$ (X = pnictide, chalcogenide, or B) based superconductors, J. Phys.: Condens. Matter {\bf 20}, 342203 (2008).

\bibitem{Bauer2008}E. D. Bauer, F. Ronning, B. L. Scott, and J. D. Thompson, Superconductivity in SrNi$_2$As$_2$ single crystals,  Phys. Rev. B {\bf 78}, 172504 (2008).

\bibitem{Simonson2012a} J. W. Simonson, Z. P. Yin, M. Pezzoli, J. Guo, J. Liu, K. Post, A. Efimenko, N. Hollmann, Z. Hu, H.-J. Lin, C.-T. Chen, C. Marques, V. Leyva, G. Smith, J. W. Lynn, L. L. Sun, G. Kotliar, D. N. Basov, L. H. Tjeng, and M. C. Aronson, From antiferromagnetic insulator to correlated metal in pressurized and doped LaMnPO, PNAS {\bf 109}, E1815 (2012).

\bibitem{Simonson2011} J. W. Simonson, K. Post, C. Marques, G. Smith, O. Khatib, D. N. Basov, and M. C. Aronson, Gap states in insulating LaMnPO$_{1-x}$F$_x$ ($x=0$--0.3), Phys. Rev. B {\bf 84}, 165129 (2011).

\bibitem{Sun2012} Y.-L. Sun, J.-K. Bao, Y.-K. Luo, C.-M/ Feng, Z.-A. Xu, and G.-H. Cao, Insulator-to-metal transition and large thermoelectric effect in La$_{1-x}$Sr$_x$MnAsO, Europhys. Lett. {\bf 98}, 17009 (2012). 

\bibitem{Hanna2013}T. Hanna, S. Matsuishi, K. Kodama, T. Otomo, S.-I. Shamoto, and H. Hosono, From antiferromagnetic insulator to ferromagnetic metal: Effects of hydrogen substitution in LaMnAsO, Phys. Rev. B {\bf 87}, 020401(R) (2013).

\bibitem{Singh2009b} Y. Singh, M. A. Green, Q. Huang, A. Kreyssig, R. J. McQueeney, D. C. Johnston, and A. I. Goldman, Magnetic order in \bma\ from neutron diffraction measurements, Phys. Rev. B {\bf 80}, 100403(R) (2009).

\bibitem{Calder2014} S. Calder, B. Saparov, H. B. Cao, J. L. Niedziela, M. D. Lumsden, A. S. Sefat, and A. D. Christianson, Magnetic structure and spin excitations in BaMn$_2$Bi$_2$, Phys. Rev. B {\bf 89}, 064417 (2014).

\bibitem{Satya2011} A. T. Satya, A. Mani, A. Arulraj, N. V. Chandra Shekar, K. Vinod, C. S. Sundar, and A. Bharathi, Pressure-induced metallization of \bma, Phys. Rev. B {\bf 84}, 180515(R) (2011).

\bibitem{Pandey2012} A. Pandey, R. S. Dhaka, J. Lamsal, Y. Lee, V. K. Anand, A. Kreyssig, T. W. Heitmann, R. J. McQueeney,  A. I. Goldman, B. N. Harmon, A. Kaminski, and D. C. Johnston, Ba$_{1-x}$K$_x$Mn$_2$As$_2$: An Antiferromagnetic Local-Moment Metal, Phys. Rev. Lett. {\bf 108}, 087005 (2012).

\bibitem{Bao2012} J.-K. Bao, H. Jiang, Y.-L. Sun, W.-H. Jiao, C.-Y. Shen, H.-J. Guo, Y. Chen, C.-M. Feng, H.-Q. Yuan, Z.-A. Xu,
G.-H. Cao, R. Sasaki, T. Tanaka, K. Matsubayashi, and Y. Uwatoko, Weakly ferromagnetic metallic state in heavily doped Ba$_{1-x}$K$_x$Mn$_2$As$_2$, Phys. Rev. B {\bf 85}, 144523 (2012).

\bibitem{Pandey2013b} A. Pandey, B. G. Ueland, S. Yeninas, A. Kreyssig, A. Sapkota, Y. Zhao, J. S. Helton, J. W. Lynn, R. J. McQueeney, Y. Furukawa, A. I. Goldman, and D. C. Johnston, Coexistence of Half-Metallic Itinerant Ferromagnetism with Local-Moment Antiferromagnetism in Ba$_{0.6}$K$_{0.4}$Mn$_2$As$_2$,  Phys. Rev. Lett. {\bf 111}, 047001 (2013).

\bibitem{Pandey2015} A. Pandey and D. C. Johnston, Ba$_{0.4}$Rb$_{0.6}$Mn$_2$As$_2$: A prototype half-metallic ferromagnet, Phys. Rev. B {\bf 92}, 174401 (2015).

\bibitem{Ueland2015} B. G. Ueland, A. Pandey, Y. Lee, A. Sapkota, Y. Choi, D. Haskel, R. A. Rosenberg, J. C. Lang, B. N. Harmon, D. C. Johnston, A. Kreyssig, and A. I. Goldman, Itinerant Ferromagnetism in the As $4p$ Conduction Band of ${\rm Ba_{0.6}K_{0.4}Mn_2As_2}$ Identified by X-ray Magnetic Circular Dichroism, Phys. Rev. Lett. {\bf 114}, 217001 (2015).

\bibitem{Lamsal2013} J. Lamsal, G. S. Tucker, T. W. Heitmann, A. Kreyssig, A. Jesche, A. Pandey, W. Tian, R. J. McQueeney, D. C. Johnston, and A. I. Goldman, Persistence of local-moment antiferromagnetic order in Ba$_{1-x}$K$_x$Mn$_2$As$_2$, Phy. Rev. B {\bf 87}, 144418 (2013).

\bibitem{Das2017} P. Das, N. S. Sangeetha, A. Pandey, Z. A. Benson, T. W. Heitmann, D. C. Johnston, A. I. Goldman, and A. Kreyssig, Collinear antiferromagnetism in trigonal \sma\ revealed by single-crystal neutron diffraction, J. Phys.: Condens. Matter {\bf 29}, 035802 (2017).

\bibitem{Bridges2009} C. A. Bridges, V. V. Krishnamurthy, S. Poulton, M. P. Paranthaman, B. C. Sales, C. Myers, and S. Bobev, Magnetic order in CaMn$_2$Sb$_2$ studied via powder neutron diffraction, J. Magn. Magn. Mater. {\bf 321}, 3653 (2009). 

\bibitem{Ratcliff2009} W. Ratcliff II, A. L. Lima Sharma, A. M. Gomes, J. L. Gonzalez, Q. Huang, and J. Singleton, The magnetic ground state of CaMn$_2$Sb$_2$,  J. Magn. Magn. Mater. {\bf 321}, 2612 (2009).

\bibitem{Simonson2012} J. W. Simonson, G. J. Smith, K. Post, M. Pezzoli, J. J. Kistner-Morris, D. E. McNally, J. E. Hassinger, C. S. Nelson, G. Kotliar, D. N. Basov, and M. C. Aronson, Magnetic and structural phase diagram of CaMn$_2$Sb$_2$, Phys. Rev. B {\bf 86}, 184430 (2012).

\bibitem{Gibson2015} Q. D. Gibson, H. Wu, T. Liang, M. N. Ali, N. P. Ong, Q. Huang, and R. J. Cava, Magnetic and electronic properties of CaMn$_2$Bi$_2$: A possible hybridization gap semiconductor, Phys. Rev. B {\bf 91}, 085128 (2015).

\bibitem{Johnston2011}  D. C. Johnston, R. J. McQueeney, B. Lake, A. Honecker, M. E. Zhitomirsky, R. Nath, Y. Furukawa, V. P. Antropov, and Y. Singh, Magnetic exchange interactions in ${\rm BaMn_2As_2}$: A case study of the $J_1$-$J_2$-$J_c$ Heisenberg model, Phys. Rev. B {\bf 84}, 094445 (2011).

\bibitem{McNally2015} D. E. McNally, J. W. Simonson, J. J. Kistner-Morris, G. J. Smith, J. E. Hassinger, L. DeBeer-Schmitt, A. I. Kolesnikov, I. A. Zaliznyak, and M. C. Aronson, CaMn$_2$Sb$_2$: Spin waves on a frustrated antiferromagnetic honeycomb lattice, Phys. Rev. B {\bf 91}, 180407 (2015).

\bibitem{Xie2017} W. Xie, M. J. Winiarski, T. Klimczuk, and R. J. Cava, A tetragonal polymorph of \smp\ made under high pressure---theory and experiment in harmony, Dalton Trans. {\bf 46}, 6835 (2017).

\bibitem{Brechtel1979} E. Brechtel, G. Cordier, and H. Sch\"afer, Preparation and Crystal Structure of \bms, BaZn$_2$Sb$_2$ and BaCd$_2$Sb$_2$, Z. Naturforsch. {\bf 34b}, 921 (1979).

\bibitem{Cordier1976} G. Cordier and H. Schaefer, New intermetallic compounds in the anti-dicerium dioxide sulfide structure type, Z. Naturforsch. {\bf 31b}, 1459 (1976).

\bibitem{Xia2008}S.-Q. Xia, C. Myers, and S. Bobev, Combined Experimental and Density Functional Theory Studies on the Crystal Structures and Magnetic Properties of Mg(Mg$_{1-x}$Mn$_x$)$_2$Sb$_2$ ($x\approx 0.25$) and \bms, Eur. J. Inorg. Chem. {\bf 27}, 4262 (2008). 

\bibitem{An2009}J. An, A. S. Sefat, D. J. Singh, and M.-H. Du, Electronic structure and magnetism in \bma\ and \bms, Phys. Rev. B {\bf 79}, 075120 (2009).

\bibitem{Bobev2006} S. Bobev, J. Merz, A. Lima, V. Fritsch, J. D. Thompson, J. L. Sarrao, M. Gillessen, and R. Dronskowski, Unusual Mn--Mn Spin Coupling in the Polar Intermetallic Compounds CaMn$_2$Sb$_2$ and \sms, Inorg. Chem. {\bf 45}, 4047 (2006).

\bibitem{fullprof} J. Rodr\'iguez-Carvajal, Physica B \textbf{192}, 55 (1993).

\bibitem{APEX2015} APEX3, Bruker AXS Inc., Madison, Wisconsin, USA, 2015.

\bibitem{SAINT2015} SAINT, Bruker AXS Inc., Madison, Wisconsin, USA, 2015.

\bibitem{Krause2015} L. Krause, R. Herbst-Irmer, G. M. Sheldrick, and D. J. Stalke, Appl. Crystallogr. {\bf 48}, 3 (2015).

\bibitem{Sheldrick2015A} G. M. Sheldrick, SHELTX---Integrated space-group and crystal-structure determination, Acta Crystallogr. A {\bf 71}, 3 (2015).

\bibitem{Sheldrick2015C} G. M. Sheldrick, Crystal structure refinement with SHELXL, Acta Crystallogr.~C {\bf 71}, 3 (2015).

\bibitem{Okita1968} T. Okita and Y. Makino, Crystal Magnetic Anisotropy and Magnetization of MnSb, J. Phys. Soc. Jpn. {\bf 25}, 120 (1968).

\bibitem{Chen1977} T. Chen, G. B. Charlan, and R. C. Keezer, Growth of MnSb Single Crystals by Puling with a Seed from Nonstoichiometric Molten Solution, J. Cryst. Growth {\bf 37}, 29 (1977).

\bibitem{Johnston2012} D. C. Johnston, Magnetic Susceptibility of Collinear and Noncollinear Heisenberg Antiferromagnets, Phys Rev. Lett. {\bf 109}, 077201 (2012).

\bibitem{Johnston2015} D. C. Johnston, Unified molecular field theory for collinear and noncollinear Heisenberg antiferromagnets, Phys. Rev. B {\bf 91}, 064427 (2015).

\bibitem{Malaman1994} B. Malaman, G. Venturini, R. Welter, and E. Ressouche, Neutron diffraction studies of CaMn$_2$Ge$_2$ and BaMn$_2$Ge$_2$ compounds: first examples of antiferromagnetic Mn planes in \tcs-type structure compounds, J. Alloys Compd. {\bf 210}, 209 (1994).

\bibitem{Goetsch2012}  R. J. Goetsch, V. K. Anand, A. Pandey, and D. C. Johnston, Structural, thermal, magnetic, and electronic transport properties of the LaNi$_2$(Ge$_{1-x}$P$_x$)$_2$ system, Phys. Rev. B {\bf 85}, 054517 (2012).

\end{thebibliography}
\end{document}